\pdfoutput=1
\documentclass[]{elsarticle}

\usepackage[utf8x]{inputenc}
\usepackage{ucs}
\usepackage{graphicx}	% Including figure files
\usepackage{amssymb} % for \geqslant and \leqslant
\usepackage{amsmath} % for \text{} in equations and 'align' environment

\usepackage[left=3cm,right=3cm,top=3cm,bottom=3cm]{geometry}

\biboptions{sort&compress}
\usepackage{har2nat} % "natbib" is loaded automatically

\usepackage{xcolor}
\definecolor{lightpink}{rgb}{1,.90,.90}
\definecolor{lightred}{rgb}{1,.50,.50}
\definecolor{lightblue}{rgb}{.80,.90,1}
\definecolor{lightorange}{rgb}{1.0,0.8,0.5}

\usepackage[bookmarks=true,colorlinks=true,citecolor=blue,linkcolor=blue,urlcolor=magenta]{hyperref}

\usepackage{soulutf8}
\begin{document}

\begin{frontmatter}

%\title{Evidence of dark matter axions identical to solar axions}
\title{Experimental evidence of dark matter axions identical to solar axions and the absence of the ``fifth'' carrier force for the Higgs field}

\author{Vitaliy~D.~Rusov\footnote{Corresponding author: Vitaliy D. Rusov, siiis@te.net.ua}}
\author{Igor~V.~Sharph}
\author{Vladimir~P.~Smolyar}
\address{Department of Theoretical and Experimental Nuclear Physics,\\Odessa Polytechnic National University, Odessa, Ukraine}

\begin{abstract}
%Based on the modified Turner model~\cite{Turner1987,Turner1986Report} in the early epochs of the Universe after inflation and before the Big Bang nucleosynthesis, we obtain an estimate of the fraction of critical density $\Omega_{axionHDM}^{thermal} \sim 0.32$ with a mass of $\sim6~eV$.
%% which contains both the thermal axions of hot dark matter with a mass of $\sim 40~eV$, and coherent axons of warm dark matter with a mass of $\sim 3.2 \cdot 10^{-2}~eV$.
%%We show that, due to the energy of the axion field, when the PQ symmetry breaking occurs after inflation, 
%Next, due to the energy of the axion field, thermal axions of hot dark matter in the dominant era of the Universe between the Big Bang nucleosynthesis and the cosmological microwave background (CMB), transform into other fields and generate particles represented as mixed dark matter as $\left[ \Omega_{axionHDM}^{thermal} \right]_{\sim 0.32} = \Omega_{ADM} + \Omega_{baryon}$ and $\left[ \Omega_{axionHDM}^{thermal} \right]_{axiogenesis} = \Omega_{axoinWDM}^{coherent} + \Omega_{baryon}$, where the ADM with a mass of 5~GeV is the result of the decay of the dark matter of the Higgs boson, whose partial composite of component is directly related to the ``visible'' composite of Higgs boson, and the second term is $\Omega_{axoinWDM}^{coherent} \sim 4.5 \cdot 10^{-5}$, where the mass of coherent axions of warm dark matter is $\sim 3.2 \cdot 10^{-2} ~eV$.
%

%\noindent \colorbox{yellow}{\parbox{\linewidth}{
The complete ``experimental'' proof of the existence of axion luminosity and trapped ADMs in the interior of the Sun is the reason for the formation of hot dark matter axions (see modified Turner model~\cite{Turner1987,Turner1986Report} with a critical density fraction of $\Omega_{axionHDM}^{thermal} \sim 0.32$ at a mass of $\sim6~eV$) between inflation and Big Bang nucleosynthesis, which are further transformed into mixed dark matter between Big Bang nucleosynthesis and the cosmological microwave background (CMB) in the form of $\left[ \Omega_{axionHDM}^{thermal} \right]_{\sim 0.32} = \Omega_{ADM} + \Omega_{baryon}$ and $\left[ \Omega_{axionHDM}^{thermal} \right]_{axiogenesis} = \Omega_{axoinWDM}^{coherent} + \Omega_{baryon}$, 
%with the first term $\left[ \Omega_{axionHDM}^{thermal} \right]_{\sim 0.32} \sim 0.32$, 
where the first term 5~GeV ADM is the result of the decay of the partially composite dark matter of the Higgs boson, in which there is a remarkable absence of the ``fifth'' carrier force for the Higgs field in the Universe, and the second term $\Omega_{axoinWDM}^{coherent} \sim 4.5 \cdot 10^{-5}$ is associated with the mass of $\sim 3.2 \cdot 10^{-2} ~eV$ of coherent axions of warm dark matter, which are identical to solar axions with the same mass~\cite{RusovDarkUniverse2021}.

\end{abstract}

%\begin{keyword}
% \sep  \sep  \sep 
%\end{keyword}

\end{frontmatter}

%\begin{keywords}
%\end{keywords}

\newlength{\myIndent}
\setlength{\myIndent}{\parindent}

\section{Introduction and motivation}

A hypothetical pseudoscalar particle called axion is predicted by the theory
related to solving the CP-invariance violation problem in QCD. The most
important parameter determining the axion properties is the energy scale $f_a$
of the so-called U(1) Peccei-Quinn symmetry violation. It determines both the
axion mass and the strength of its coupling to fermions and gauge bosons
including photons. However, in spite of the numerous direct experiments, axions
have not been discovered so far. Meanwhile, these experiments together with the
astrophysical and cosmological limitations leave a rather narrow band for the
permissible parameters of invisible axion (e.g.
$10^{-6} eV \leqslant m_a \leqslant 10^{-2} eV$~\citep{ref01,ref02}).
The PQ mechanism, solving the strong CP problem in a very elegant 
way~\citep{PecceiQuinn1977,PecceiQuinn1977PRD,Wilczek1978,Weinberg1978}, is
especially attractive here, since the axion is also a candidate for
dark matter~\citep{Preskill1983,Abbott1983,Dine1983,Kawasaki2013,Marsh2016,DiLuzio2020,Sikivie2021}.

\begin{figure}[tb!]
\noindent
  \begin{center}
    \includegraphics[width=16cm]{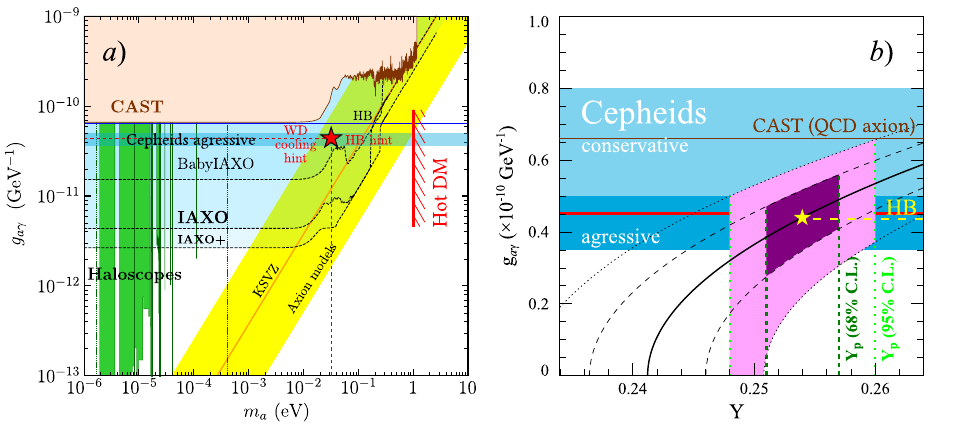}
  \end{center}
\caption{(a) Summary of astrophysical, cosmological and laboratory constraints
on axions. Comprehensive axion parameter space, highlighting two main front
lines of direct detection experiments: helioscopes (CAST~\cite{CAST2017}) and
haloscopes (ADMX~\cite{Asztalos2010}, and RBF~\cite{Wuensch1989} and microwave resonators~\cite{Brubaker2017,Zhong2018,Braine2020,Lee2020,Backes2021}).
% \hl{Lehnert \& Carosi 2017 ?}).
The astrophysical bounds from horizontal branch and massive stars are labeled
``HB''~\cite{Raffelt2008} and ``Cepheids''~\cite{Carosi2013}, respectively, and
there are also astrophysical hints (WD cooling hints, and HB hint). 
The QCD motivated models (KSVZ~\cite{Kim1979,Shifman1980}) for axions lay in
the yellow diagonal band. A plot of $g_{a\gamma}$ versus $m_a$ with the most
stringent results (solid lines) and sensitivity perspectives (dashed lines) of
observations and experiments directly comparable to the different phases of
IAXO are shown, BabyIAXO, IAXO, and an upgraded version of IAXO, IAXO+. 
The yellow band denotes the region of the parameter space favoured by QCD axion
models. The red star marks the values of the axion mass 
$m_a \sim 3.2 \cdot 10^{-2}~eV$ and the axion-photon coupling constant 
$g_{a\gamma} \sim 4.4\cdot 10^{-11}~GeV^{-1}$, which
were first obtained experimentally (see~\cite{RusovDarkUniverse2021}). \\
(b) $R$ parameter constraints, which compares the numbers of stars in the the
horizontal branch (HB) and in the upper portion of the red giant branch (RGB),
to helium mass fraction $Y$ and axion coupling $g_{a\gamma}$ (adopted 
from~\cite{Ayala2014}). The resulting bound on the axion 
($g_{10} = g_{a\gamma} / (10^{-10}~GeV^{-1})$ is somewhere between rather 
conservative $0.5 < g_{10} < 0.8$ and most aggressive 
$0.35 < g_{10} < 0.5$~\cite{Friedland2013}. 
The red line marks the value of the axion–photon coupling constant 
$g_{a\gamma} \sim 4.5\cdot 10^{-11}~GeV^{-1}$ adopted from~Eq.~(2) 
in~\cite{Ayala2014}. The blue shaded area represents the bounds from Cepheids
observation. The yellow star corresponds to $Y=0.254$ and the bounds from HB
lifetime (yellow dashed line).}
\label{fig-axion-constraints}
\end{figure}

By now, we have shown that axions born in the core of the Sun, and photons of axion origin (in the solar corona with a temperature of one million degrees) lead to a strict limit on the axion-photon coupling constant $g_{a\gamma} \sim 4.4\cdot 10^{-11}~GeV^{-1}$ and mass $m_a \sim 3.2 \cdot 10^{-2}~eV$ (see Fig.~\ref{fig-axion-constraints}a; also Fig.1 in~\cite{RusovDarkUniverse2021}). Most importantly, hadronic axions originating from the core of the Sun are controlled by 11-year variations in the density of asymmetric dark matter (ADM), which is gravitationally trapped in the interior of the Sun (through vertical density waves from the black hole disk to the neighborhood of the Sun). The hard part of the spectrum of solar photons is of axion origin, and is defined as the product of the fraction of axions in the solar core and the fraction of the area of sunspots. The latter is the result of variations in the solar luminosity (see Eqs.~(27)-(28) in~\cite{RusovDarkUniverse2021}). This means that solar luminosity variations and other solar cycles are the result of 11-year ADM density variations that are anticorrelated with the density of hadronic axions in the Sun's core.

Interestingly, in contrast to our Fig.~\ref{fig-axion-constraints}a, in Fig.~\ref{fig-axion-constraints}b, one can see a well-known strict limitation on the axion-photon coupling $g_{a\gamma} \sim 4.5\cdot 10^{-11}~GeV^{-1}$, but in a wide mass range (see Eq.~(2) in~\cite{Ayala2014}) from an analysis of a sample of 39 Galactic Globular Clusters. However, in our case (Fig.~\ref{fig-axion-constraints}a) the wide range of masses in~\cite{Ayala2014} is replaced by a single mass $m_a \sim 3.2 \cdot 10^{-2}~eV$, which fits the axions of galactic globular clusters~\cite{Ayala2014}, but also all axion stars in our galaxy, including neutron stars~\cite{Buschmann2021}, supernovae~\cite{Carenza2019,Carenza2021}, and the Sun~\cite{RusovDarkUniverse2021}.

\begin{figure}[tb!]
  \begin{center}
  \includegraphics[width=15cm]{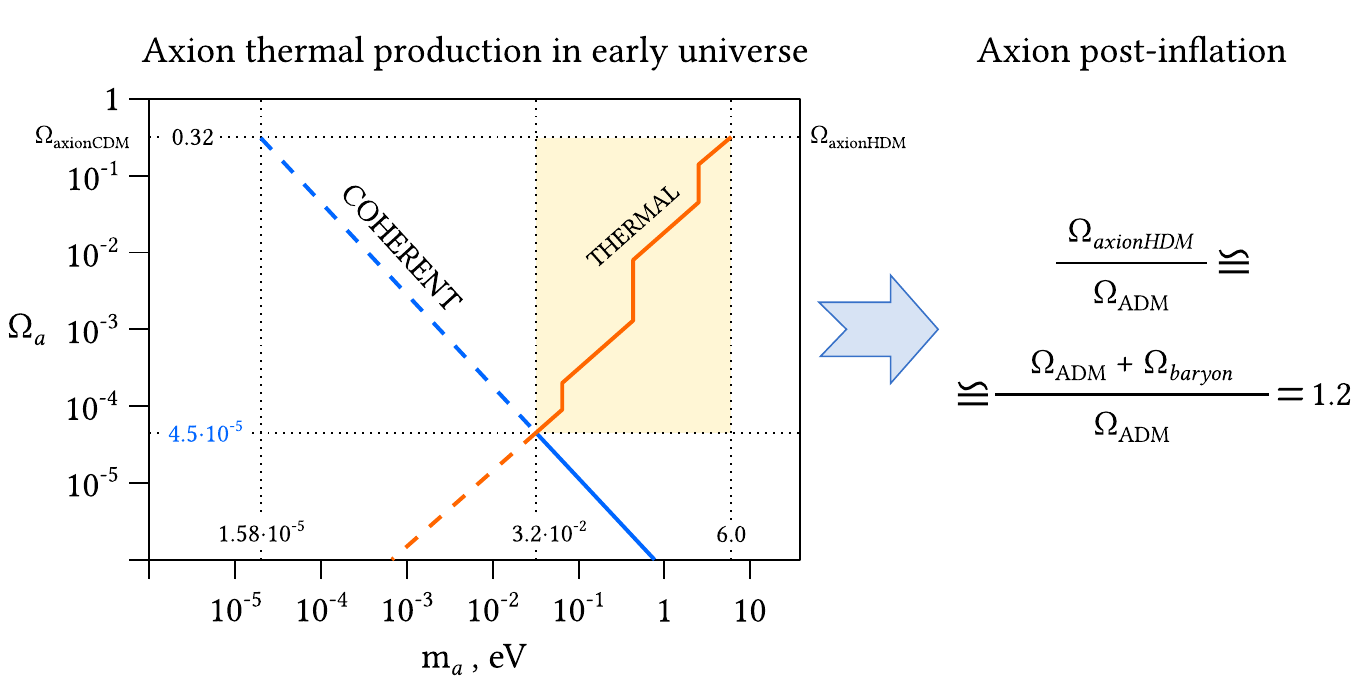}
  \end{center}
\caption{Thermal and coherent axion contributions to the fraction of 
critical density ($\rho_c = 3 H_0^2 / 8\pi G = 1.05 \cdot 10^4 h^2 eV/cm^3$) as
a function of axion mass (left panel adapted from~\cite{Turner1986Report}). 
On the one hand, $m_a \sim 3.2\cdot 10^{-2}~eV$ establishes the identity of the
thermal and coherent axions ($\Omega_{thermal} / \Omega_{coherent} \sim 1$), 
which simultaneously obey the relation 
$\Omega_{coherent} = \Omega_{thermal} \sim 4.5 \cdot 10^{-5}$ based on the
coherent value of the relative critical density of the Universe 
$\Omega_a \approx (6\mu eV / m_a)^{7/6}$~\cite{Graham2015}.
And on the other hand, two final populations of axions appear simultaneously --
the thermal with a mass of $\sim 6.0 ~eV$ (see~\cite{PDG2020} for comparison)
and $\Omega_{axionHDM} \sim 0.32$ (see left panel), and the coherent with a 
mass of $m_a \sim 3.2\cdot 10^{-2}~eV$ and 
$\Omega_{axionWDM}^{coherent} \approx 4.5 \cdot 10^{-5}$.
We also know~\cite{RusovDarkUniverse2021} that ADM with 
$\Omega_{DM} \cong 5.3 \Omega_{baryon}$~\cite{PlanckCol2015,Patrignani2016} and
energy density $\Omega_{axionHDM}^{post-inflation} / \Omega_{ADM} \cong 1.2$ 
(see right panel) is formed at the initial stage of the post-inflation 
Universe, and still exists.}
\label{fig-axion-inflation}
\end{figure}

We know that, according to the modified Turner model~\cite{Turner1987,Turner1986Report} (see also Fig.~1 in~\cite{RusovDarkUniverse2021}), the 
Peccei-Quinn symmetry breaking scale is $\sim 2\cdot 10^8~GeV$
(corresponding to the axion mass of $\sim 3.2 \cdot 10^{-2} ~eV$ (see our
Fig.~\ref{fig-axion-constraints}, and also Fig.~1 
in~\cite{RusovDarkUniverse2021})), with which there is a thermal production of hot dark
matter axions\footnote{In our case, the limits on the axion mass are
established by cosmology and astrophysics: axions are members of cold dark
matter when the mass (and energy scale) is at $\sim 2 \cdot 10^{-5}~eV$ (and 
$f \sim 3 \cdot 10^{11}~GeV$) or hot dark matter when the mass (and energy scale)
is at $40~eV$ (and $f \sim 1.5 \cdot 10^{5}~GeV$). Between cold and hot dark matter
there is a warm dark matter with a mass of $\sim 3.2 \cdot 10^{-2}~eV$ (and 
$f \sim 1.9 \cdot 10^8~GeV$) (see left panel in Fig.~\ref{fig-axion-inflation} from the modified Turner model~\cite{Turner1987,Turner1986Report}). A more general representation of the dependence of the mass
$m_a$ and the energy scale $f$ is given in Fig.~1 in~\cite{Graham2015}.}
in the Universe (via Primakoff processes ($q  + \gamma \rightarrow q + a$; 
the particle in the loop is the heavy $Q$ quark) and photoproduction on heavy
quarks (see paragraphs above Eqs.~(4) and (5) in~\cite{Turner1987}; and Fig.~1 in~\cite{Turner1986Report}).
According to the Turner model~\cite{Turner1987}, dark matter axions in the 
early Universe contain two populations of axions -- thermal (orange) and 
coherent (blue) (see left panel in Fig.~\ref{fig-axion-inflation}),
which are both thermal axions of hot dark matter with a mass of 6.0~eV and $\Omega_{axionHDM}^{thermal} \sim 0.32$ and coherent axions of warm dark matter with a mass of $m_a \sim 3.2 \cdot 10^{-2}~eV$ and $\Omega_{axionWDM}^{coherent} \approx 4.5 \cdot 10^{-5}$ (see the left panel in Fig.~\ref{fig-axion-inflation}). With this axion-photon-photon bond, the axion decays into two photons ($axion \rightarrow 2 \gamma$) with a lifetime (see e.g.~\cite{Moroi1998})

\begin{align}
\tau_a & = \left[ \frac{\alpha C_{a\gamma\gamma}^2}{256 \pi^2} \cdot \frac{m_a^3}{f_a^2} \right]^{-1} \cong
1.2 \cdot 10^{12} yr C_{a\gamma\gamma}^{-2} \left( \frac{m_a}{10 eV} \right)^{-5} \approx  \nonumber \\ 
& \approx 1.2 \cdot 10^{12} yr \times 12.86 \left( \frac{m_a}{6} \right)^{-5} = 1.5 \cdot 10^{13} yr \left( \frac{m_a}{6 eV} \right)^{-5} \approx  \nonumber \\
& \approx 10^4 ~billion~yr .
\label{eq-PLB7-01}
\end{align}

Notice that the lifetime of the axion is longer (see Eq.~(\ref{eq-PLB7-01}1): $\tau_a \approx 10^4$ billion years for $m \sim 6~eV$ and $C_{a\gamma\gamma} \leqslant 1$) than the age of the Universe of $\sim13.77$ billion years (see e.g. Table~1 in~\cite{PlanckCol2018} and Table~4 in~\cite{Aiola2020}), and hence primordial axions are still in the early Universe. However, as we will see later, radiative decay of the axion may affect the background UV photons in spite of the long lifetime.

In order to solve the fundamental problem of the identity of dark matter axions
and solar axions, we must first consider the physics of the Turner 
model~\cite{Turner1987,Turner1986Report} (see also Fig.1 
in~\cite{RusovDarkUniverse2021}), within which the relative critical density of
the Universe with the curve $\Omega_a - m_a$ completely does not coincide with
$\Omega_{coherent} = \Omega_{thermal} \sim 4.5 \cdot 10^{-5}$ and mass
$3.2\cdot 10^{-2}~eV$ at the point of intersection between the coherent and 
thermal axions (see left panel in Fig.~\ref{fig-axion-inflation}), although it
is known that the presence of a mass of $3.2\cdot 10^{-2}~eV$ is completely
determined by the known coherent value of the Universe critical density
fraction $\Omega_a \approx (6\mu eV / m_a)^{7/6}$~\cite{Graham2015}.

Here we use the energy density of the Universe today, at which the fraction of
the critical density $\Omega_a$ (see Eq.~(7) in~\cite{Turner1987}, which is
exactly the same as Eq.~(10) in~\cite{Turner1986Report}) contributed by axions
with masses $m_a$ and their number density today $n_a$, as well as the critical
density $\rho_c$ with the participation of the scale factor of the Hubble
expansion rate with the so-called $h$, has a contribution in the form

\begin{equation}
\Omega_a = \rho_a / \rho_c = n_a \times m_a / \rho_c \approx 
n_a \times m_a / (1.05 \cdot 10^4 h^2 eV / cm^3) ,
\label{eq-PLB7-02}
\end{equation}

\noindent where

\begin{equation}
n_a = 13 \times (60 / g_{*d}) \times T_{2.7}^3 cm^{-3} .
\label{eq-PLB7-03}
\end{equation}

Here $T_{2.7}$ is the current microwave temperature in units of 2.7~K and 
$h \sim 0.6736$, e.g.~\cite{PlanckCol2018}.

From here, according to the Turner model (see Eq.~(7) in~\cite{Turner1987} or Eq.~(10) in~\cite{Turner1986Report}, accurately describing Fig.~3 equations (10) in~\cite{Turner1986Report}), the final result of the energy density of the Universe today will be

\begin{equation}
\Omega_{thermal} \cong (T_{1/2}^3 / h^2) \times 1.3 \cdot 10^{-3} \times m_{eV} \times (60/g_{*d}) \approx \frac{2.9 \times (m_a / eV)}{10^3} \times \frac{60}{g_{*d}} .
\label{eq-PLB7-04}
\end{equation}

Here $g_{*d} \equiv g_{*} (T_d)$, where $g_*$, as usual, counts the effective number of relativistic degrees of freedom at the epoch when the $a$ particles thermally decouple in the early Universe. In this case, according to Turner (see the first sentence after Eq.~(7) in~\cite{Turner1987}), for a temperature $T_d$ (near thermal equilibrium (see Fig.~2 in~\cite{Turner1986Report})), which leads to about 1-100~GeV, there appears $g_{*d} \approx 60$, while for $T_d \leqslant 0.1~GeV$, $g_{*d} \approx 10$.

Oddly enough, this raises an unexpected question: Why doesn't the exact Eq.~(\ref{eq-PLB7-03}) (see Eq.~(7) in~\cite{Turner1987} and Eq.~(10) in~\cite{Turner1986Report}) for the relative critical density of the Universe give the correct result for $\Omega_{thermal}$?! In other words, why can we never get the right kink in the curve (see $60/g_{*d} \sim 1 \div 0.49$ with $m_a \approx 0.07~eV$; and also the left panel in Fig.~\ref{fig-axion-inflation}) for thermal production results, when $g_{*d}$ only changes from about $g_{*d} \approx 60$ to about $g_{*d} \approx 10$ (see the first sentence after Eq.~(7) in~\cite{Turner1987}) for $m_a \approx 2.6~eV$, which is why $\Omega_a \approx 10^{-4}$ for $m_a \approx 3.2 \cdot 10^{-2}~eV$ (see left panel in Fig.~\ref{fig-axion-inflation}) becomes an order of magnitude larger (unfortunately!) than the one that must obey the known coherent value of the fraction of the critical density of the Universe $\Omega_a \approx 4.5 \cdot 10^{-5}$ for $m_a \approx 3.2 \cdot 10^{-2}~eV$ as $\Omega_a \approx (6\mu eV /m_a)^{7/6}$~\cite{Graham2015}.

As a result, we realized that instead of the old known Turner numbers (see Eq.~(\ref{eq-PLB7-04})) such as $g_{*d} \approx 60$ at $T_d \sim 1-100~GeV$ and $g_{*d} \approx 10$ at $T_d \leqslant 1~MeV$, we now use more accurate up to date numbers of $g_{*d} \sim 120$ at $T_d \sim 100~GeV$ (see the first sentence after Eq.~(\ref{eq-PLB7-04})), $g_{*d} \approx 60$ at $T_d \sim 10~GeV$ and $g_{*d} \approx 10$ at $T_d \leqslant 1~MeV$  and $g_{*d} \sim 3.24$ at $T_d \sim 10~keV$, related to the nature of the Big Bang Nucleosynthesis (BBN) temperature from 1~MeV to 10~keV (see~\cite{DeSimone2019}). The importance of these calculations (see Eq.~(\ref{eq-PLB7-04}) and left panel in Fig.~\ref{fig-axion-inflation}) is due to the fact that we knew the value of the relative critical density of the Universe with the $\Omega_a$ -- $m_a$ curve and a mass of $3.2 \cdot 10^{-2}~eV$ in advance, which exactly coincided with the intersection points between coherent and thermal axions (see left panel in Fig.~\ref{fig-axion-inflation}). This means that $\Omega_{axionWDM}^{coherent} = \Omega_{axionHDM}^{thermal} \approx 4.5 \cdot 10^{-5}$ with a mass of $3.2 \cdot 10^{-2}~eV$, at which the Peccei-Quinn symmetry breaking scale is $f_a \sim 2 \cdot 10^8 ~GeV$ (see modified Turner's model (Figs.~\ref{fig-axion-constraints},~\ref{fig-axion-inflation} and~\cite{Turner1987,Turner1986Report})) is determined by means of $\Omega_a \approx (6\mu eV /m_a)^{7/6}$~\cite{Graham2015}.

It is known that
%, according to the work of Co~\&~Harigaya~\cite{Co2020}, 
%we use a
%mechanism called axiogenesis, in which the cosmological excess of baryons over
%antibaryons is generated from the rotation of QCD axion. This is because
the PQ symmetry is an approximate global symmetry, which is clearly broken by the QCD
anomaly. Given that the symmetry is not exact (since quantum gravity does not
allow global symmetry~\citep{Giddings1988,Coleman1988,Gilbert1989,Harlow2019,Harlow2019arxiv,Co2020}),
it is likely that the PQ symmetry was significantly broken in the early Universe
and caused the rotation of QCD axion~\cite{Co2020}, whose cosmological role
generates an excess of baryons over antibaryons. It is also noted that the
baryon asymmetry is transformed by QCD and electroweak sphaleron transitions.

\begin{figure}[tb!]
  \begin{center}
  \includegraphics[width=14cm]{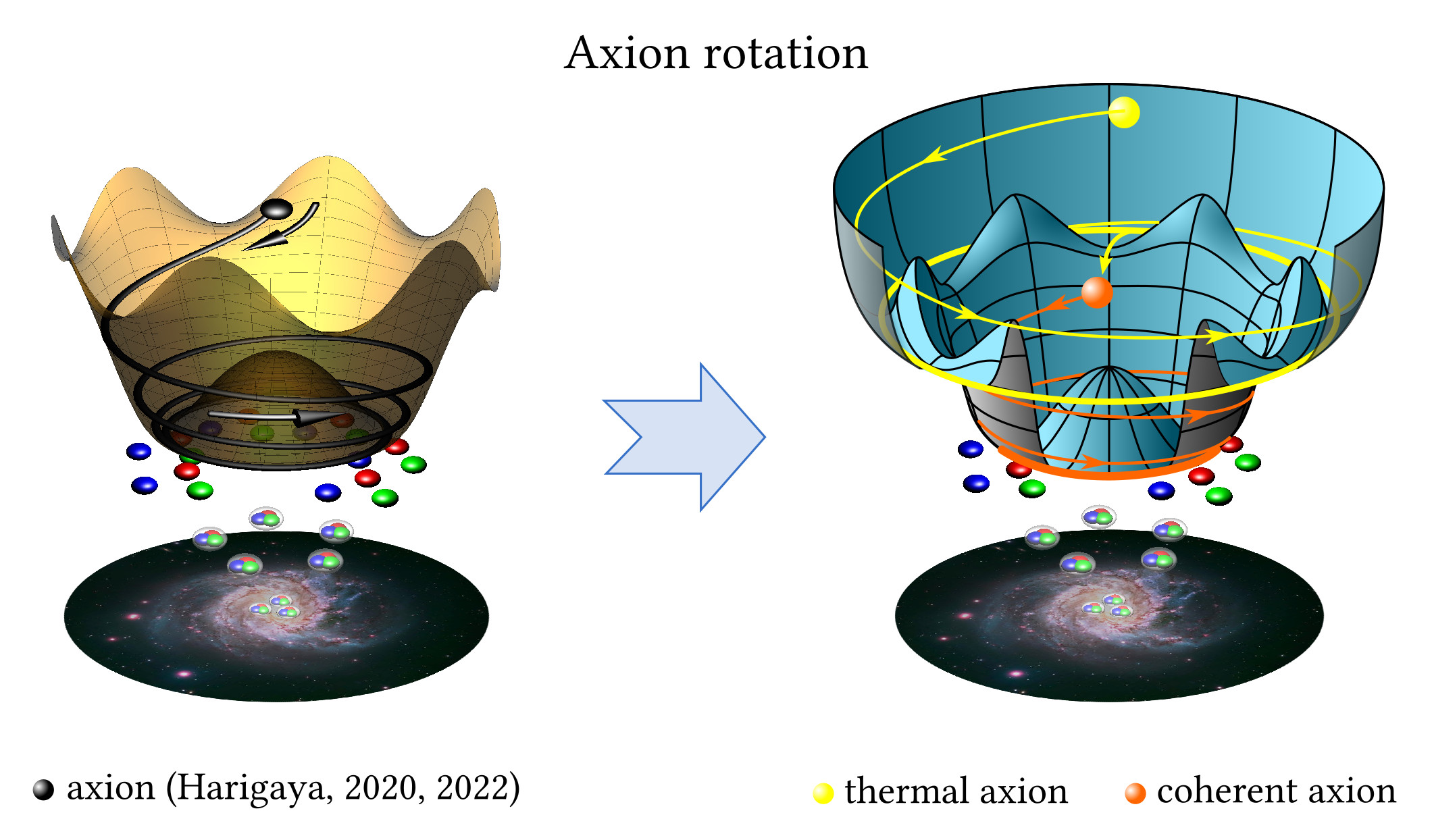}
  \end{center}
\caption{\textbf{Left:} Mexican hat-like potential of the Peccei-Quinn field at
temperatures above (below) the critical temperature of the QCD phase 
transition~\cite{Harigaya2020Conf,Harigaya2022Conf,Harigaya2023UCLAConf}. At high temperatures, when
the kinetic energy of rotation, exceeding the height of axion potential (left 
panel from~\cite{Harigaya2023UCLAConf}), generates a
high concentration of ``invisible'' axions, or more precisely, the hot dark
matter axion in the post-inflation. As a result, the kinetic energy of the
axion field flows into other fields and generates particles that appear in a
form of excess of matter over antimatter in the Universe (see left panel).
At low temperatures of the axion potential, when the axion ``marble'' settles,
it will revolve around this minimum (left panel 
from~\cite{Harigaya2023UCLAConf}), generating coherent axion 
of warm dark matter, for which the Peccei-Quinn symmetry scale energy is
$\sim 2 \cdot 10^8 ~GeV$ (see modified Turner's model, Figs.~\ref{fig-axion-constraints},~\ref{fig-axion-inflation} and~\cite{Turner1987,Turner1986Report}).\\
\textbf{Right:} 
To clarify the understanding of the essence of the figure on the left, we
supplemented it with our right figure, reflecting the existence of the speed of
the ``yellow'' thermal HDM axion with a mass of$m_a \sim 6~eV$ (or 
$f_a \sim 10^6 ~GeV$; see left panel in Fig.~\ref{fig-axion-constraints}). 
When this axion ``breaks through'' into the axion field after some base of
kinetic displacement (see~\cite{Co2020Misalignment}), it appears as an 
``orange'' coherent axion. In other words, the speed of the ``yellow'' thermal
HDM axion is converted to the speed of the ``orange'' coherent HDM axion (see
analog of ``black'' axion in left panel). When the ``angular'' velocity first
rotates above the top of the potential (see Figure~1 kinetic displacement
mechanism in~\cite{Co2020Misalignment}), then the interaction of the 
``orange'' coherent HDM axion (see the ``black'' axion analog in the left
panel) with the quarks results in the formation of particles that are generated
in the form of asymmetric baryons (see left panel and right panel). On the 
other hand, when the ``angular'' velocity of the coherent HDM axion at the base
of the potential (see Misalignment Mechanism in Fig.~1 
in~\cite{Co2020Misalignment}) becomes zero, $\dot{\theta} = 0$, then the 
``radial'' velocity of the ``orange'' thermal HDM axion is the source of birth 
of coherent axion WDM.}
\label{fig-rotating-axion}
\end{figure}

But the most remarkable fact is how efficient the transfer rate from the rotation of the axion of hot dark matter to quarks and the asymmetry of matter and antimatter is. When the kinetic energy of rotation becomes higher than the potential barrier of the axion, it is transformed into a large number of axions, and thanks to the kinetic energy of the axion field, these axions of hot dark matter are transferred into other fields and generate particles that appear as the excess of matter over the antimatter in the Universe (see left panel in Fig.~\ref{fig-rotating-axion}). When the kinetic energy of rotation falls below the potential of the axion, a large number of ``invisible'' axions practically disappear due to the formation of the asymmetry of matter, and the remaining axion ``ball'' settles down to the point of the minimum of the axion field (see left panel in Fig.~\ref{fig-rotating-axion}), generating an axion with low density of warm DM (WDM).

But here appears the main and surprising result. The
radial velocity of the ``yellow'' thermal axion of HDM (see right side in 
Fig.~\mbox{\ref{fig-rotating-axion}}), at which the angular 
velocity $\dot{\theta} = 0$ appears at the base of the potential (see 
Misalignment Mechanism in Fig.~1 in~\cite{Co2020Misalignment}), is the source of the coherent axions of WDM only (see right panel in Fig.~\ref{fig-rotating-axion}). On the other hand, when the angular velocity manifests itself only up to the top of the potential (see Kinetic Misalignment Mechanism of Fig.~1 in~\cite{Co2020Misalignment}), it is the source of the formation of asymmetric baryons only, the number of which is 5 times less than the formation of a coherent axion of WDM. From here, we understand that axion abundance becomes enhanced~\cite{Co2020Misalignment} compared to the traditional misalignment mechanism~\cite{Preskill1983,Abbott1983,Dine1983}. According to axiogenesis (see Eq.~5 in~\cite{Co2020}), an abundance of thermal axions of hot dark matter 

\begin{align}
\frac{\Omega_a h^2}{\Omega_{DM} h^2} & \approx 140 \left( \frac{f_a}{10^8 ~GeV} \right) \left( \frac{130 ~GeV}{T_{WS}} \right)^2 \left( \frac{0.1}{c_B} \right) \equiv \nonumber \\
& \equiv \frac{\left[ \Omega_{axionHDM}^{thermal} h^2 \right]_{axiogenesis}}{\Omega_{axionWDM}^{coherent} h^2} = \frac{\Omega_{axionWDM}^{coherent} + \Omega_{baryon}}{\Omega_{axionWDM}^{coherent}} \cong 1.2
\label{eq-PLB7-05}
\end{align}

\noindent
must be much larger for $f_a \sim 10^6 ~GeV$ (or $m_a \sim 6~eV$ and $\left[ \Omega_{axionHDM}^{thermal} \right]_{axiogenesis} = 1.2 \times \Omega_{axionWDM} \sim 4.5 \cdot 10^{-5}$; see left panel in Fig.~\ref{fig-axion-constraints}) than the observed DM abundance $\Omega_{DM} h^2$ (or $\left[ \Omega_{axionHDM}^{thermal} \right]_{axiogenesis} / 1.2 = \Omega_{axionWDM} \sim 4.5 \cdot 10^{-5}$) for $f_a \sim 2 \cdot 10^8 ~GeV$ (or $m_a \sim 3.2 \cdot 10^{-2} ~eV$), satisfying astrophysical constraints~\cite{Ellis1987,Raffelt1988SN1987A,Turner1988,Mayle1988,Raffelt2008,Payez2015,Bar2020}, the prediction of $T_{WS} \cong 130~GeV$~\cite{DOnofrio2014}, and $c_B = O(0.117)$. A value of $f_a \sim 10^6 ~GeV$ leads to both successful axiogenesis and axion dark matter and interestingly resides in the so-called hadronic axion window~\cite{Turner1988,Chang1993,Moroi1998}, which however is recently under scrutiny (see e.g.~\cite{Carenza2019}).

Here two very interesting points in the modified Turner model come into play (see~\cite{Turner1987,Turner1986Report} and also Fig.~1 in~\cite{RusovDarkUniverse2021}), where the Peccei-Quinn symmetry breaking scale is $\sim 2\cdot 10^8 ~GeV$, which corresponds to an axion mass of $\sim 3.2 \cdot 10^{-2} ~eV$ (see Fig.~\ref{fig-axion-constraints}, and also Fig.~1 in~\cite{RusovDarkUniverse2021}). Note that $m_a \sim 3.2 \cdot 10^{-2} ~eV$ sets the identity of the thermal and coherent axions, $\Omega_{thermal} / \Omega_{coherent} \sim 1$ (see left panel in Fig.~\ref{fig-axion-inflation}), which simultaneously obey $\Omega_{coherent} = \Omega_{thermal} \approx 4.5 \cdot 10^{-5}$. In the dominant epoch of the Universe, at very high temperatures between inflation and the Big Bang of nucleosynthesis, the ``coherent-thermal'' axions of warm dark matter transform into thermal axions of hot dark matter with a mass of $\sim 6 ~eV$ (and $\Omega_{axionHDM} \sim 0.32$). When the temperature drops during nucleosynthesis and further after nucleosynthesis to the cosmological microwave background, surprisingly, thermal axions of hot dark matter with a mass of $\sim 6 ~eV$ will transform back into coherent axions of warm dark matter with a mass of $\sim 3.2 \cdot 10^{-2} ~eV$ (and $\Omega_{axionWDM} \sim 4.5 \cdot 10^{-5}$). 

From here, the first point is in what way the ``coherent-thermal'' axions of warm dark matter are identical to solar axions (see Section~\ref{sec-axions-identity-proof}). The second point is why thermal axions of hot dark matter transform exactly into ADM, and why ADM is driven by baryonic symmetry (see Section~\ref{sec-ADM-Higgs-proof}).

The main connection between two very interesting points is due to the fact that 11-year variations of the solar axion are the reason of the anticorrelated 11-year variations in the ADM density in the interior of the Sun, predetermined by vertical density waves from the disk (or black hole) to the solar neighbourhood~\cite{RusovDarkUniverse2021}. 
At the same time, although relic thermal axions of hot dark matter with a mass $m_a \sim 6.0 ~eV$ are practically inaccessible for direct cosmological searches, they are available for direct astrophysical experiments as additional participants in mixed dark matter -- relic coherent axions warm dark matter with a mass of $m_a \sim 3.2 \cdot 10^{-2} ~eV$ identical to solar axions (see~\cite{RusovDarkUniverse2021}) and ADM particles with a mass of $m_{ADM} \sim 5~GeV$ caused by the dark matter decay of the Higgs boson (see Section~\ref{sec-ADM-Higgs-proof}), in which it is theoretically and experimentally shown that that the fifth carrier force of the Higgs field does not exist!

Our paper is organized as follows. In Section~\ref{sec-mixedDM} we get a new physics of mixed dark matter: axion rotations and the ADM phenomenon controllable by the baryon asymmetry of the Universe. In Section~\ref{sec-axions-identity-proof}, we discuss the proof of the identity of dark matter axions in the Universe. In Section~\ref{sec-ADM-Higgs-proof}, we discuss evidence for the formation of ADM in the Universe caused by the decay of the dark matter Higgs boson, and the absence of the ``fifth'' carrier force for the Higgs field. In Section~\ref{sec-Higgs-experiment}, we obtain experimental proof of the existence in the Universe of both dark matter axions and ADM caused by the decay of dark matter Higgs boson. We are completing with Section~\ref{sec-conclusions}.

\section{The mixed dark matter: axion rotations and the phenomenon of ADM controlled by the baryon asymmetry of the Universe}
\label{sec-mixedDM}

The new mixed dark matter physics leads to the following description of the generalized model (directly related to the modified Turner model (see the analogue of $\Omega_{axionHDM}$ in~\cite{Turner1987,Turner1986Report}) and our model with $\Omega_{ADM}$ due to Higgs dark matter decay~\cite{Merkotan2017,Merkotan2018,Ptashynskiy2019,Merkotan2021PhD}))

\begin{equation}
1 - \Omega_{DE} \equiv \Omega_{axionHDM}^{thermal} = \left[ \Omega_{axionHDM}^{thermal} \right]_{\sim 0.32} 
+ \left[ \Omega_{axionHDM}^{thermal} \right] _{axiogenesis} \cong 0.32 ,
\label{eq-PLB7-06}
\end{equation}

\noindent
where $\Omega_{DE}$ is the dark energy (see e.g.~\cite{Li2020,Li2019} [45,46]), and dark matter contains

\begin{equation}
\frac{\left[ \Omega_{axionHDM}^{thermal} \right]_{\sim 0.32}}{\Omega_{ADM}} =
\frac{\Omega_{ADM} + \Omega_{baryon}}{\Omega_{ADM}} = 1.2 ,
\label{eq-PLB7-07}
\end{equation}

\begin{equation}
\frac{\left[ \Omega_{axionHDM}^{thermal} \right]_{axiogenesis}}{\Omega_{axionWDM}^{coherent}} =
\frac{\Omega_{axionWDM}^{coherent} + \Omega_{baryon}}{\Omega_{axionWDM}^{coherent}} = 1.2 ,
\label{eq-PLB7-08}
\end{equation}

\noindent
where Eq.~(\ref{eq-PLB7-07}) matches the right panel in Fig.~\ref{fig-axion-inflation}, Eq.~(\ref{eq-PLB7-08}) matches Eq.~(\ref{eq-PLB7-05}) and $\Omega_{ADM} > \Omega_{baryon} \gg \Omega_{axionWDM}^{coherent} (= 4.5 \cdot 10^{-5})$ (see Eq.~(\ref{eq-PLB7-09})).

In contrast to equation~(\ref{eq-PLB7-08}), which solves the beautiful problem of transforming the abundance of thermal axions of hot dark matter as $\left[ \Omega_{axionHDM}^{thermal} \right]_{axiogenesis} / 1.2 \cong \Omega_{axionWDM}^{coherent} \sim 4.5 \cdot 10^{-5}$ with mass $m_a \sim 6~eV$ (or $f_a \sim 10^6 ~GeV$; see Eq.~(\ref{eq-PLB7-05}) and left panel in Fig.~\ref{fig-rotating-axion}) into asymmetric baryons and coherent axions of the WDM with a mass of $\sim 3.2 \cdot 10^{-2} ~eV$ (or $f_a \sim 2 \cdot 10^8 ~GeV$; see Eq.~(\ref{eq-PLB7-05}) and left panel in Fig.~\ref{fig-rotating-axion}), there is a significant problem in equation~(\ref{eq-PLB7-07}), which should somehow solve the problem of $\left[ \Omega_{axionHDM}^{thermal} \right]_{\sim 0.32}$, and a single explanation of baryon and dark matter densities through the formation of asymmetric dark matter from baryogenesis. This problem, although ``simple'' in principle, can be very complex in practice.

In particular, this applies to such significant problems as the identity of dark matter axions and solar axions, and the physics of ADM formation caused by the Higgs dark matter decay.

\subsection{Proof for the identity of dark matter axions in the Universe and the solar axions}
\label{sec-axions-identity-proof}

The main fundamental problem of mixed dark matter is the following. When the dominant era of the universe is between the end of inflation and the beginning of the Big Bang nucleosynthesis, then the formation of abundances of thermal axions of hot dark matter is observed ($\Omega_{axionHDM}^{thermal}$ with a mass $\sim 6~eV$ or $f_a \sim 10^6~GeV$), for which we previously used the Peccei-Quinn symmetry breaking scale with axion decay constant $f_a \sim 2 \cdot 10^8~GeV$ (see~\cite{Turner1987,Turner1986Report}) and axion mass $\sim 3.2 \cdot 10^{-2} ~eV$ (see Fig.~\ref{fig-axion-constraints}, as well as Fig.~1 in~\cite{RusovDarkUniverse2021}). After the onset of the epoch of Big Bang nucleosynthesis and before the cosmological microwave background (CBM), the abundance of hot dark matter thermal axions transforms into $\left[ \Omega_{axionHDM}^{thermal} \right]_{\sim 0.32} = \Omega_{ADM} + \Omega_{baryon}$ (see Eq.~(\ref{eq-PLB7-07})) and $\left[ \Omega_{axionHDM}^{thermal} \right]_{axiogenesis} = \Omega_{axionWDM}^{coherent} + \Omega_{baryon}$ (see Eq.~(\ref{eq-PLB7-08})). In this case, ADM particles with a mass of $\sim 5~GeV$ (at $\Omega_{ADM} \sim 0.27$) and coherent axions of WDM with a mass of $\sim 3.2 \cdot 10^{-2} ~eV$ (at $\Omega_{axionWDM}^{coherent} \sim 4.5 \cdot 10^{-5}$) should exist in the form of non-clustered ADM particles and clustered axions~\cite{Turner1987,Turner1986Report}, which are associated with primary density perturbations in the Universe.

If the structure formation in the Universe occurs because of the primary density perturbations that grow due to Jeans instability\footnote{The first attempts to investigate the existence of fluid instability in the context of collapsing astrophysical bulges were initiated by Sir Jeans in 1902~\cite{Jeans1902,Jeans1928}. According to~\cite{Yang2020}, in the their seminal work, Jeans first described quantitatively the collapse of the matter system due to its self-gravitational interaction based on the ideal hydrodynamics model, and found that under the effect of a small perturbation in the forms of $\exp (-i\omega t + ikx)$ with the dispersion equation $\omega ^2 = k^2 c_S^2 - \Omega _G ^2$, where $\Omega _G = \sqrt{4\pi G \rho}$ is the Jeans gravitational frequency, $\rho$ is the matter density, $G$ is the gravitational constant, and $c_S$ is the adiabatic sound velocity, a self-gravitational infinite uniform gas in a static state should be unstable. From the dispersion equation, a noteworthy conclusion can be found, that is, when the wave-number $k$ is less than the Jeans wave-number $k_J \equiv \sqrt{\Omega_G / c_S}$ the value of $\omega^2$ is negative and Jeans instability increases.} (see~\cite{Kremer2016,Kremer2018,Kremer2019,Yang2020}), then axions and ADM should participate in structure formation~\cite{Turner1987,Turner1986Report}. We are interested in the physics of interstellar gas clouds, which have been considered ideal sites for star formation due to the gravitational collapse of the Jeans instability (see~\cite{Snell1981,Dwivedi1999,Kremer2018,Kremer2019}). There is strong evidence that the mass component of the Universe is mainly composed of baryons, dark matter and dark energy against the background of standard models in both particle physics and cosmology~\cite{Snell1981,Kremer2018,Kremer2019}. Therefore, as key ingredients of star formation, both dark matter and baryons can exist together in interstellar gas clouds. It is known that the critical Jeans mass, above which an interstellar gas cloud undergoes such a gravitational collapse, depends on the temperature and density of the cloud. In this case, the compression of the cloud of interstellar matter, caused by gravitational collapse, raises the temperature until thermonuclear fusion occurs in the center of the star and typically ranges from thousands to tens of thousands of solar masses~\cite{Prialnik2009}.

\begin{figure}[tb!]
  \begin{center}
  \includegraphics[width=14cm]{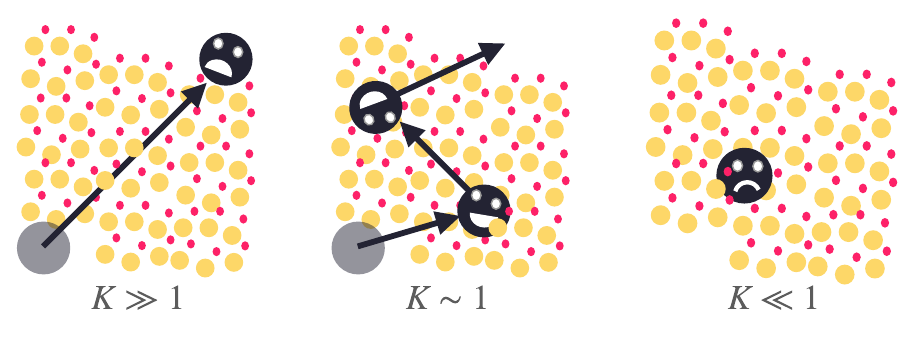}
  \end{center}
\caption{Three heat conduction regimes by dark matter in the solar plasma (figure from~\cite{Vincent2020}; see three Knudsen regimes in~\cite{Dearborn1990}). Left (Knudsen regime): large mean free paths (small $\sigma$) computable using the Spergel and Press approach~\cite{Spergel1985}, lead to low overall energy deposition. Right (LTE regime): small mean free paths, computable with the Gould and Raffelt approach~\cite{GouldRaffelt1990a,GouldRaffelt1990b}, mean the DM gets ``stuck'' as $\sigma$ grows. Center: at the Knudsen transition, heat transport is optimized. This regime does not have an analytical solution and must be calculated with a Monte Carlo-calibrated interpolation. The SP solution is based on incorrect assumptions, and GR can be numerically unstable and breaks down at small radii.}
\label{fig-smiling-dark-matter}
\end{figure}

At relatively ``not high'' temperatures and density of the interstellar gas cloud, the grouped axions ``act'' mainly in the form of so-called three-particle axion absorptions\footnote{This was first pointed out by Anselm and Uraltsev~\cite{Anselm1982}.} $a + e^{-} + (e^{-}, Ze) \rightarrow e^{-} + (e^{-}, Ze)$. They ``accumulate energy'' of absorbed axions in the cores of stars, in particular in the core of a very young Sun. This roughly coincides with another ``conducting energy'' of a relatively small amount of asymmetric dark matter (ADM with a density of $\sim 0.38 ~GeV/cm^3$) through the ADM-nucleon interactions. It can act as a heat conductor (see Fig.~\ref{fig-smiling-dark-matter} for the Knudsen transition at the $K \sim 1$ from~\cite{Vincent2020}, which is in the middle of three Knudsen regimes in~\cite{Dearborn1990}) by acquiring kinetic energy from the hot core, and releasing it in interactions with nuclei in the cooler outer regions. This could result in a slightly smaller temperature gradient in the star.

When the compression of a cloud of interstellar matter, caused by gravitational collapse, leads to an increase in temperature and density and, as a result, to the occurrence of thermonuclear processes in the centers of stars, in particular, in the core of a very young Sun, it generates both the  two-particle ``rebirth'' of axions ($\gamma + (e^{-}, Ze) \rightarrow (e^{-}, Ze) + a$~\cite{Raffelt1986}), and as always, ADM particles through the ADM-nucleon coupling, which suppress the number of ``reborn'' axions of warm dark matter.

\begin{figure}[tb!]
  \begin{center}
  \includegraphics[width=10cm]{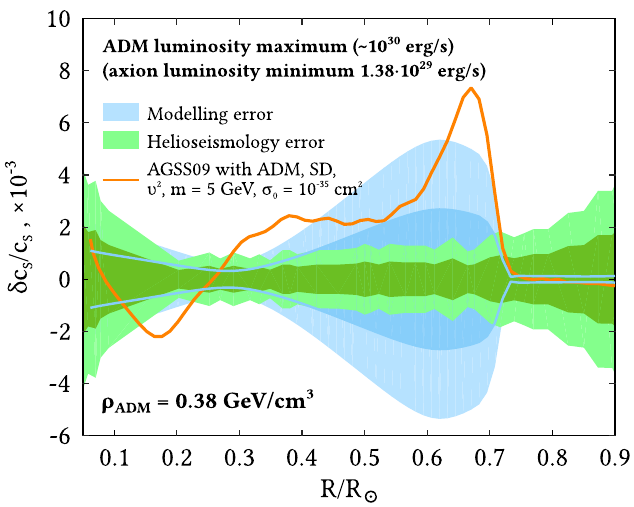}
  \end{center}
\caption{Deviation of the radial speed of the sound profile $c_s(r)$ between the standard solar models with the capture and transfer of heat of asymmetric dark matter (solid orange line ADM~\cite{Vincent2016}) and with axions of warm dark matter~\cite{Grevesse1998}.
The dark and light blue bands represent the 1 and $2\sigma$ errors in modelling, estimated from error propagation of uncertainties in SSM inputs (see analogous~\cite{Serenelli2013,Villante2014,Vinyoles2015}). The green bands represent the 1 and $2 \sigma$ errors in helioseismological inversions~\cite{DeglInnoccenti1997,Fiorentini2001} [24, 24]. The ADM particles, using the ADM-nucleon coupling, on the one hand, suppress the number of ``reborn'' axions of warm dark matter in the core of the Sun, and on the other hand -- suggest the complete absence of 11-year variations of the radial sound speed profile $\delta c_s/c_s$ in the interior of the Sun, since the local density of dark matter in the solar neighborhood is practically constant!}
\label{fig-helioseismology-no-variations}
\end{figure}

From here we understand that here arises a very interesting and intriguing problem of the Sun. It is known that at present there are 11-year variations of the ADM density inside the Sun, which controls the inverse variations of the solar luminosity of axions (see Eqs.~(27)-(28) in~\cite{RusovDarkUniverse2021}). When we decrease the age of the Sun ($\sim 4570 ~Myr$~\cite{Connelly2012}), we are interested in the age of the Sun, which is smaller than before $\leqslant 20 ~Myr$ backward~\cite{Subr2019}, but greater than or equal to the age of the epoch of nucleosynthesis. This means that at that time there were no S-stars (see Fig.~8 in~\cite{RusovDarkUniverse2021}) between the black hole and the disk in our galaxy and, as a result, there were no 11-year ADM density variations in the galaxy halo and on the Sun (see Fig.~\ref{fig-helioseismology-no-variations}).

In the case when $L_a^{max} \gg L_{ADM}^{min}$, we are interested in the age of the Sun, which depends both on the ``beginning'' of the age ($4570~Myr - 20~Myr = 4550~Myr$) and up to the present time ($\sim 4570~Myr$~\cite{Connelly2012}. It means there are already S-stars between the black hole and the disk in our galaxy (see Fig.~8 in~\cite{RusovDarkUniverse2021}) and, as a result, the 11-year variations of the ADM density appear in the galaxy halo, and particularly, on the Sun (see Fig.~\ref{fig-helioseismology} in section~\ref{sec-Higgs-experiment}). At the same time, it should be remembered~\cite{RusovDarkUniverse2021} that the variations of axion luminosity on the Sun and other solar cycles, in particular, sunspots, are the result of 11-year variations of the ADM density, which are anticorrelated with the density of dark matter axions in the Sun's core.

From here, we recall that the so-called three-particle absorptions of axions $a + e^{-} + (e^{-}, Ze) \rightarrow e^{-} + (e^{-}, Ze)$, which ``accumulate energy'' of absorbed axions in the cores of stars, leads to an increase in the temperature and density of a highly compressed cloud of interstellar matter, caused by gravitational collapse, and, as a consequence, to the occurrence of thermonuclear processes in the centers of stars, in particular, in the core of a very young Sun. There, a two-particle ``rebirth'' of axions happens ($\gamma + (e^{-}, Ze) \rightarrow (e^{-}, Ze) + a$), but, due to $L_a^{max} \gg L_{ADM}^{min}$ (see e.g. Fig.~\ref{fig-helioseismology}a), without the participation of ADM particles with ADM-nucleon coupling. At the same time, axion streams are radiated from the core to the surface of the Sun and further into the space of the Galaxy\footnote{It is amusing that the solar conditions for the three-particle absorption of axions give the same mean free path as the two-particle absorption, $a + (e^{-}, Ze) \rightarrow (e^{-}, Ze) + \gamma$~\cite{Goulding2018}. With an axion mass of about $3.2 \cdot 10^{-2} ~eV$, it is about $10^{23} ~cm$. This is a little more than the radius of the Milky Way galaxy. Therefore, stars can be considered transparent from the point of view of axion propagation.}. 
%\hl{Then at high temperatures and densities of clouds, in which thermonuclear processes are manifested in the centers of stars, there are two-particle production of axions $\gamma + (e^{-}, Ze) \rightarrow (e^{-}, Ze) + a$ (different in translation). In this case, axion fluxes are emitted from the core to the surface of the Sun and further into the space of the Galaxy.} 
Eventually, it can be said that during and after the epoch of Big Bang nucleosynthesis, ``collisions and interactions between gas-rich galaxies are considered key stages in their formation and evolution, causing the rapid formation of new stars''~\cite{Goulding2018}, and, as a consequence, the emergence of the radiating streams of coherent axions of warm dark matter from their cores into outer space (for example, neutron stars~\cite{Buschmann2021}, white dwarf cooling~\cite{Rusov2015}, supernovae~\cite{Carenza2019,Carenza2021}, galactic globular clusters~\cite{Ayala2014}, the Sun~\cite{RusovDarkUniverse2021}).

As a result, new physics is manifested not only in Eq.~(\ref{eq-PLB7-10}), but also in the description of the low density (with $\Omega_{axionWDM}^{coherent} = 4.5 \cdot 10^{-5} ~eV$) of coherent axions of warm dark matter in interstellar gas clouds in the Universe, creating a hierarchy of condensed structures such as clusters of galaxies, stellar groups, stars and planets, in particular, on the Sun (see Eq.~(17) in~\cite{Graham2015})

\begin{equation}
\Omega_{axionWDM}^{coherent} = \Omega_{\lbrace solar \rbrace axion} \approx
\left( \frac{6 \mu eV}{m_a} \right)^{7/6} = 4.5 \cdot 10^{-5} ,
\label{eq-PLB7-09}
\end{equation}

\noindent
which, on the basis of ``experimental'' data, identically coincides with the values of the hadronic axion-photon coupling $g_{a\gamma} \sim 4.40 \cdot 10^{-11} ~GeV^{-1}$ and the mass $m_a \sim 3.2 \cdot 10^{-2} ~eV$ on the Sun (see Fig.~\ref{fig-axion-constraints} and Fig.~1 in~\cite{RusovDarkUniverse2021}). This means that the solar axions and the warm dark matter axion remnants in the ADM halo (see Eq.~(\ref{eq-PLB7-10})) are absolutely identical!

Below we show the experimental evidence of the identity of solar axions and warm dark matter axion remnants in the galactic halo ADM.

\subsection{Proof for the formation of ADM in the Universe caused by the decay of dark matter Higgs boson}
\label{sec-ADM-Higgs-proof}

The main idea of the ADM formation is as follows. According to our works~\cite{Volkotrub2015,Chudak2016,Merkotan2017,Merkotan2018,Ptashynskiy2019,Merkotan2021PhD}, it can be shown that
Higgs vector fields or Higgs tensor fields with interior indices are the result of
the known ``visible'' Higgs boson (see e.g. $m_H = 125.8 \pm 0.14 ~GeV$ from~\cite{CMS2020}), while the scalars in the interior indices can be considered as components of the dark matter Higgs boson. This means that when an abundance of hot dark matter thermal axions generate a baryon asymmetry, the latter contain not only the ``visible'' Higgs bosons (see~\cite{CMS2022a,CMS2020,CMS2022b}), but also the dark matter components of the Higgs boson (see~\cite{Merkotan2017,Merkotan2018,Merkotan2021PhD}. Moreover, according to our assumption, it is the decay of the dark matter Higgs boson that leads to the formation of dark matter fermions as the asymmetric dark matter with a mass of $\sim 5~GeV$ (see e.g.~\cite{Vincent2016} and Eqs.~(24)-(25) in~\cite{RusovDarkUniverse2021}), at which the well-known ``cosmic coincidence'' does not appear by chance, since the relict abundance of asymmetric dark matter is determined (also not by chance!) by means of the baryon asymmetry in the Universe!
%\hl{This is not proof, but after some close time we will finally get the proof of a unified explanation for the asymmetry densities of baryons and dark matter and send our article to the journal. (different in translation)}

When the fractions of the critical densities of DMs and baryons with $\Omega_{DM} = 5.3 \Omega_{baryon}$ are close to unity~\cite{Planck2015,Patrignani2016}, we no longer have the question ``why the cosmological ``coincidence problem'' is being discussed now'' (see, for example,~\cite{Velten2014}). In contrast to the well-known answers to this question~\cite{Nussinov1985,Chivukula1990,Barr1990,Kaplan1992,Hooper2005,Kaplan2009}, there is a view of asymmetric dark matter (ADM) (for recent reviews see~\cite{Davoudiasl2012,Petraki2013,Zurek2014}) in which the DM density is determined by the asymmetry $\eta_{DM}$ in the same way as the baryon asymmetry number $\eta_{baryon}$ sets the density of the visible matter. Since we already assume that one asymmetry $\eta_{DM}$ is controlled by another asymmetry $\eta_{baryon}$, we expect $\eta_{DM} \sim \eta_{baryon}$. 
%This is not a proof, but a suggestion of the ADM formation through the decay of the \hl{dark matter Higgs boson, which is a component of the visual Higgs boson}(\cite{Merkotan2017,Merkotan2018,Merkotan2021PhD}; let us note that the Higgs lifetime is close to $\tau_H \sim 1.6 \cdot 10^{-22} ~s$). Hence, abundance is associated with
Using the assumption of the formation of ADM through the decay of the dark matter of the Higgs boson, which is the component of the visual composite Higgs boson, we get the relation

\begin{equation}
\frac{\Omega_{DM}}{\Omega_{baryon}} \approx 5.3 \approx \frac{n_{DM}}{n_{baryon}} \frac{m_{DM}}{m_N} \approx \frac{m_{DM}}{m_N} ,
\label{eq-PLB7-10}
\end{equation}

\noindent
where $m_{DM} = m_{ADM} \approx 5~GeV$ is the ADM mass, which matches very well with ``experimental'' data in Section~\ref{sec-Higgs-experiment} (see left panel in our Fig.~\ref{fig-helioseismology}, and also Fig.~6 in~\cite{Vincent2016}), and $m_N \approx 0.939~GeV$ is the nucleon mass.

We know that ADM provides a natural explanation for the comparable densities of baryonic and dark matter. This is due to the fact that ADM, as a participant of the ADM halo in solar neighborhood (see~\cite{RusovDarkUniverse2021}), almost exactly coincides with the experimental data in our Section~\ref{sec-Higgs-experiment}. At the same time, the experimental hint says that although the physics of this problem is quite complicated, but, surprisingly, very simple in principle, as ADM is ``born'' (through the decay of the dark matter of the Higgs boson) from the baryon asymmetry of the Universe!

This means that, using the values for ADMs, and according to the physics of axiogenesis (see Eq.~(\ref{eq-PLB7-07}), also Eq.~5 in~\cite{Co2020}),
% values previously obtained from the decay of \hl{cold (different in translation)} dark matter Higgs (see~\hl{\mbox{\cite{Merkotan2018}}}), 
we get

\begin{equation}
\frac{\left[ \Omega_{axionHDM}^{thermal} \right]_{\sim 0.32}}{\Omega_{ADM}} =
\frac{\Omega_{ADM} + \Omega_{baryon}}{\Omega_{ADM}} = 1 + m_N / m_{ADM} \sim 1.2 ,
\label{eq-PLB7-11}
\end{equation}

\noindent
which predetermines an apparent identity between $\Omega_{ADM} \sim 0.27$ and $\Omega_{baryon} \sim 0.05$ (see right panel in Fig.~\ref{fig-axion-inflation} and  Eqs.~(\ref{eq-PLB7-07}) and~(\ref{eq-PLB7-11})).

Surprisingly, this raises two more fundamental problems. At the beginning, we are interested in the first unique fundamental problem. We know that the Standard Model of particle physics has difficulties related to the fact that a number of important components of this model do not have experimental confirmation. In particular, the U(1) symmetry of the Lagrangian of the Standard Model is supported by a gauge field, whose quanta, like the quanta of one of the components of the SU(2) gauge field, are not observed in the experiment, but their linear combinations are observed instead -- photons and Z-bosons. The Standard Model does not explain why some states are observed and others are not. Secondly, the spontaneous symmetry breaking mechanism (see e.g.~\cite{Beekman2019,CERNHiggs2022,CMS2022a,Grojean2007}) associated with the postulation of the non-gauge Higgs self-coupling, which together with the Yukawa coupling between fermionic fields and the Higgs field are two new interactions that cannot be reduced to any of the four known interactions, and the existence of which has no experimental confirmation. Thirdly, the Higgs boson of the Standard Model is a particle with a weak isospin of 1/2, while its decay modes, known from experiments, show that the final decay state is characterized by the integer weak isospin. These circumstances predetermine a completely new modification of the Standard Model. 

As noted by Salam~\textit{et~al.}~\cite{Salam2022}, ``Physicists’ current theory of fundamental particles and forces is known as the Standard Model, a theoretical framework that provides a description of elementary particles and the forces that make them interact with one another, with the exception of gravity. Within the Standard Model, the Higgs field is essential to describe the world as we know it.'' Surprisingly, we know him, unlike most physicists, starting with the work of Englert and Braut~\cite{Englert1964}, Higgs~\cite{Higgs1964b} and Guralnik, Hagen and Kibble~\cite{Guralnik1964} followed by Weinberg~\cite{Weinberg1967} and Salam~\cite{Salam1964} who extended the work of Glashow~\cite{Glashow1961} and proposed using the Brout-Englert-Higgs mechanism for a theory of unification of the electromagnetic and weak interactions, called the electroweak interaction, and finally, e.g. with Tanabashi~\textit{et~al.}~\cite{TanabashiPDG2018}, Zyla~\textit{et~al.}~\cite{PDG2020}, Salam~\textit{et~al.}~\cite{Salam2022}, ATLAS Collaboration~\cite{ATLAS2022}, CMS Collaboration~\cite{CMS2022a,CMS2022b}, which, based on the existence of the Brout-Englert-Higgs mechanism, directly implies the existence of a ``fifth force'' mediated by the Higgs boson. 
And the key difference of our approach is that, first ($\bullet$) we will demonstrate the essence of the multi-particle fields, then ($\bullet\bullet$) we will consider the Higgs field as a multi-particle field, and consequently, the appearance of the composite dark matter of the Higgs boson, and finally ($\bullet\bullet\bullet$), we will show why the composite nature of the Higgs boson is necessary.

($\bullet$) \textbf{The essence of multi-particle fields.}
Probably for the first time, the idea of multi-particle fields was proposed by Yukawa~\cite{Yukawa1949,Yukawa1950a,Yukawa1950b}. Yukawa called these fields ``nonlocal'' fields. We use another term -- ``multi-particle fields'' to show the differences between our model and the model proposed by Yukawa. Our model is also essentially different from the models on the light cone~\cite{Dirac1949,Heinzl2001}, quasipotential models~\cite{Logunov1963a,Logunov1963b,Faustov1973}, and models with multitime probability amplitudes~\cite{Tomonaga1946,Dirac1932,Petrat2014a}.
And the most important difference, in our view, is that the internal variables of such fields in different inertial reference frames cannot be related to each other, whereas these variables are connected by Lorentz transformations in the mentioned models. Our view has been partially explained in our previous work~\cite{Chudak2016UJP}. The use of multitime probability amplitudes in~\cite{Tomonaga1946,Dirac1932,Petrat2014a,Marx1972,Sazdjian1986,Petrat2014b}, and other works of this direction, as well as the above-mentioned works, contradicts the principles of quantum theory, because, in our opinion, they do not consider the measuring instrument influence on the state of a microsystem properly. We explain it in more detail in~\cite{Ptashynskiy2019arXiv,Ptashynskiy2019}, where we proposed an alternative approach to ensuring the simultaneity of quantum mechanical measurements in different reference frames, and introduce a subset of simultaneity of the Cartesian product of several Minkowski spaces. On the other hand, the existing field theories are considered in such a way that all interaction effects are reduced to changes in the occupation numbers of the single-particle states of free particles. This leads to the fact that, in such models, when the dynamics of processes is described, the sum of energy-momenta of these one-particle states is conserved. Manwhile, the energy-momentum of hadrons, but not of constituent particles, must be conserved for the processes with hadrons. The model of multi-particle fields on the subset of simultaneity proposed in theh cited paper allows us to build a dynamic description which is free of such problems.

We propose a model describing the scattering of hadrons as bound states of their constituent quarks. We build the dynamic equations for the multi-particle fields on the subset of simultaneity~\cite{Ptashynskiy2019arXiv,Ptashynskiy2019}, using the Lagrange method, similarly to the case of ``usual'' single-particle fields. We then consider the gauge fields restoring the local internal symmetry on the subset of simultaneity. Since the multi-particle fields, which describe mesons as bound states of a quark and an antiquark, are two-index tensors relative to the local gauge group, it is possible to consider a model with two different gauge fields, each one associated with its own index. Such fields would be transformed by the same laws during a local gauge transformation and satisfy the same dynamic equations, but with different boundary conditions. The dynamic equations for the multi-particle gauge fields describe such phenomena as the confinement and the asymptotic freedom of colored objects under certain boundary conditions, and the spontaneous symmetry breaking under anothers. With these dynamic equations, we are able to describe the quark confinement in hadrons within a single model and their interaction during the hadron scattering through the exchange of the bound states of gluons -- the glueballs.

Though we did not consider the quantization procedure for multi-particle fields in this work, it is not different from the procedure described in our works~\cite{Chudak2016,Volkotrub2015}. The operators obtained after the quantization will describe the processes of creation and annihilation of glueballs, as shown in~\cite{Chudak2016,Volkotrub2015}. Accordingly, the considered meson field quantization leads to the operators of creation and annihilation of the mesons. The meson interaction due to the interaction of constituent quarks can be described as the exchange by scalar glueballs. This approach differs from the one-particle field approach, because, in our model, the energy-momentum conservation law holds true for hadrons as whole, and not for the constituent particles.

($\bullet\bullet$) \textbf{Multiparticle fields and the Higgs mechanism.}
\textbf{First}, we consider the Higgs field as a many-particle field (see~\cite{Merkotan2017,Merkotan2018,Merkotan2021PhD} and, accordingly, the Higgs boson as a bound state of two gauge bosons, on which the vector-tensor or scalar representation of the internal symmetry group is realized (see~\cite{Merkotan2017,Merkotan2018,Merkotan2021PhD}). Such consideration, in comparison with the single-particle field of the Standard Model, has two advantages. The major one is that, in contrast to the Standard Model~\cite{Yukawa1949,Yukawa1950a,Yukawa1950b}, there is no need to introduce a non-gauge self-action of the Higgs field in order to achieve spontaneous symmetry breaking. An essential feature of the proposed model is that the Higgs field self-action, which provides a non-zero vacuum value, is not considered as an independent non-gauge interaction, but as a manifestation of the self-action of a non-abelian gauge field and, does not require the introduction of any new interaction. Namely, since the Higgs boson is regarded as a bound state of $W^+$ and $W^-$ bosons , the interaction between Higgs bosons is a consequence of a non-Abelian weak SU(2) interaction of gauge bosons. This is formally seen from the derivation of equations for a two-particle gauge field in Section~3 and Eq.~(69) in~\cite{Merkotan2017} or Section~2 and Eq.~(18) in~\cite{Merkotan2018}. This is substantially different from other models (e.g. the well-known technicolor models), and requires neither new particles nor new interactions, the existence of which has no experimental confirmation, as compared to the Standard Model (see~~\cite{Merkotan2017,Merkotan2018,Merkotan2021PhD}).

\textbf{Second}, it has all the theoretical properties of the Higgs boson. For example, in the model of multi-particle fields, the spontaneous symmetry breaking is \textit{not} postulated as in the Standard Model, but is obtained from the dynamic equations of the model, and has a nonzero vacuum value (see Eqs.~(18)-(44) in~\cite{Merkotan2018} or Eqs.~(69)-(95-96) in~\cite{Merkotan2017}). This is derived from the equations for the two-particle gauge field (see Section~2 in~\cite{Merkotan2018} or Section~2 in~\cite{Merkotan2017}), which, apart from the vector and tensor representation of the internal symmetry group, also contain a scalar part. Therefore, it is natural to call such a field scalar by internal indices -- the Higgs field. However, this field cannot interact with the gauge field and provide mass to its components this way. These results lead to the fact that, unlike the four fundamental forces -- strong, electro-weak and gravitational interactions, there is no need to associate the Higgs field with some fifth (???) carrier force, which fills the entire Universe and is responsible for mass transfer to other elementary particles. In simple terms, this means that the concept of the Higgs field, giving mass to other elementary particles, although beautiful, cannot be in our Universe.

\textbf{Third}, in contrast to the (incomplete) Standard Model of particle physics, the Higgs boson here is supposed to have an integer weak isospin, which is consistent with the experimental data on its decay channels~\cite{Patrignani2016}. It is considered as a bound state of $W^+$ and $W^-$ bosons -- the particles with weak isospin of 1. Since the weak isospin of gauge bosons is equal to unity, the bound state can have a weak isospin of 2, 1 or 0, with three terms in the expansion into irreducible tensors in the form of Eq.~(61) in~\cite{Merkotan2017} or Eq.~(10) in~\cite{Merkotan2018}. The decay channel into two photons~\cite{Patrignani2016,PDG2020} suggests that the Higgs boson can be in a state with weak isospin 0. Decay channels into two particles with weak isospins of 1/2, for example, an electron and a positron, add an opportunity to observe the Higgs boson in a state with weak isospin 1. And the four lepton channels give a value of 2. Consequently, known decay channels~\cite{Patrignani2016,TanabashiPDG2018,PDG2020} indicate that the Higgs boson is an inappropriate state for a weak isospin, but only its integer values of 0, 1 or 2 can be obtained when measured. This property is reproduced by the Eq.~(61) in~\cite{Merkotan2017} or Eq.~(10) in~\cite{Merkotan2018}. This means that the state with zero isospin obviously cannot participate in weak interactions. It has zero electric charge and does not consist of strongly interacting particles. Therefore, such a state can interact only gravitationally. At the same time, through the energy-momentum tensor, such a scalar field can make its contribution to the gravitational field. Thus, such Higgs bosons, which are scalar in interior indices, can be considered as candidates for composite dark matter~\cite{Merkotan2017,Merkotan2018,Merkotan2021PhD}. 

($\bullet\bullet\bullet$) \textbf{Why the composite nature of the Higgs boson is necessary.}
We know that the Higgs boson is considered as a bound state of $W^+$ and $W^-$ bosons -- particles with weak isospin 1, which is consistent with experimental data on its decay channels~\cite{Patrignani2016}. This means that the Higgs boson is no longer an elementary particle, but a composite object. On the other hand, in addition to the ``visible'' composite Higgs boson, we already know that there is dark matter of the Higgs boson, which is a partially composite object of the relative ``visible'' Higgs boson.

A large number of ADM particles, which manifest themselves through the decays of the partially composite dark matter of the Higgs boson, are involved in the formation of the Universe's halo in the dominant epoch between Big Bang nucleosynthesis and the CMB (mentioned in abstract). From here, as the key ingredients of star formation, we use both mixed dark matter (cold ADM particles with a mass of $\sim 5~GeV$, and warm dark matter axions with a mass of $3.2 \cdot 10^{-2}~eV$) and baryons, which can exist together in interstellar gas clouds. When a cloud of interstellar matter compresses due to gravitational collapse, temperature and density increase until thermonuclear fusion occurs in the center of a star, for example, in the Sun. 

At present, the most important thing is that warm dark matter axions with a mass of $3.2 \cdot 10^{-2}~eV$, emerging from the core of the Sun, are controlled by anticorrelated 11 year density variations of asymmetric dark matter (ADM), which pass through vertical density waves from the disk of a black hole to the solar neighborhood, and are then gravitationally captured in the Sun interior (see~\cite{RusovDarkUniverse2021}).

And finally, in Section~\ref{sec-Higgs-experiment} we get a complete ``experimental'' proof of the existence of, first, ($\bullet$) the 11-year variations of solar dark matter axions (see Eq.~(\ref{eq-PLB7-12})-(\ref{eq-PLB7-13}) and also Eqs.~(27)-(28) in~\cite{RusovDarkUniverse2021}), which are associated with the luminosities of axions and captured ADMs in the Sun interior, and, as a consequence, 11-year variations in the solar luminosity at the temperature maximum $T_{max} \sim 4\cdot 10^6~K$ and at the temperature minimum $T_{min} \sim 1.5 \cdot 10^6~K$ (see our Fig.~\ref{fig-luminosity-temperature} and Fig.~2d in~\cite{RusovDarkUniverse2021}); and second, ($\bullet\bullet$) 11-year variations of sound speed deviation inside the Sun (see our Fig.~\ref{fig-helioseismology-no-variations}), which are associated with two interrelated 5.5-year patterns: the maximum luminosity of axions (Fig.~\ref{fig-helioseismology}a) and the maximum luminosity of the ADM (Fig.~\ref{fig-helioseismology}b).

%From here, the most important result appears, which lies in the fact that it is the complete ``experimental'' evidence associated with the luminosities of axions and captured ADMs in the  Sun's interior that, oddly enough, \hl{is a participant (complete experimental evidence ... is a participant?)} in the ADM halo in galaxies, and, as a result, stars \hl{with the participation of axions (participant ... with participation?)} of warm dark matter, in particular, on the Sun.
%
%In simple terms, this means that, on the one hand, it is the ADM (with mass of $\sim 5~GeV$) that is predetermined by the decay of the Higgs dark matter, and on the other hand, it is the axions of warm dark matter (with mass $3.2 \cdot 10^{-2}~eV$) that are identical to the solar axions!

This implies the most important result, which is that this complete ``experimental'' proof of the existence of axion luminosity and captured ADMs in the interior of the Sun is the reason for the formation of mixed dark matter in post-inflation as $\left[ \Omega^{thermal}_{axionHDM} \right]_{\sim 0.32} = \Omega_{ADM} + \Omega_{baryon}$ and $\left[ \Omega^{thermal}_{axionHDM} \right]_{axiogenesis} = \Omega_{axionWDM}^{coherent} + \Omega_{baryon}$, while the first term is $\left[ \Omega^{thermal}_{axionHDM} \right]_{\sim 0.32} \sim 0.32$, where the ADM with a mass of 5~GeV is the result of the decay of the partially composite dark matter of the Higgs boson, in which there is remarkably absence of the ``fifth'' carrier force for the Higgs field in the Universe, and the second term $\Omega_{axionWDM}^{coherent} \sim 4.5 \cdot 10^{-5}$ is related to the mass of coherent axions of warm dark matter $3.2 \cdot 10^{-2}~eV$, that are identical to the solar axions!

\section{Experimental proof of the existence of both the dark matter axions and ADM caused by the decay of dark matter of Higgs boson in the Universe}
\label{sec-Higgs-experiment}

Oddly enough, thinking about the deepest problem of the fundamental physics of
the Sun and the role of the solar axions and ADM in it, we kind of recall the
long forgotten, but physically deep question posed by Robert~H.~Dicke: ``Is there a
hidden chronometer in the depths of the 
Sun?''~\cite{Dicke1978,Dicke1979,Dicke1988}, and then we ask the today's 
question: ``Who controls the Sun or who measures the exact 'clock' inside the Sun?''.

%
%According to our work~\cite{RusovDarkUniverse2021}, a unique result of our
%model is the fact that the periods, velocities, and density modulations of
%S-stars around a black hole (more precisely, supermassive black hole, SMBH) are
%the fundamental indicator of the ADM (or more correctly, Higgs CDM) halo
%density modulation at the Galaxy Center, which closely correlates with the
%density modulation of the baryon matter near the SMBH. If the density
%modulations of the Higgs CDM halo at the GC lead to modulations of the ADM
%density on the surface of the Sun (through vertical density waves from the
%galactic disk to the solar neighborhood), then there is an ``experimental''
%anticorrelation connection between the modulation of the Higgs CDM density in
%the solar interior and the number of sunspots. Therefore, this is also true for
%the relationship between the periods of S-star cycles and the sunspot cycles
%(see ``experimental'' anticorrelated data in Fig.~25 
%in~\cite{RusovDarkUniverse2021}).

According to our work~\cite{RusovDarkUniverse2021}, a unique result of our
model is the fact that the periods, velocities, and density modulations of
S-stars around a black hole (more precisely, supermassive black hole, SMBH) are
the fundamental indicator of the ADM halo
density modulation at the Galaxy Center, which closely correlates with the
density modulation of the baryon matter near the SMBH. If the density
modulations of the ADM halo at the GC lead to modulations of the ADM
density on the surface of the Sun (through vertical density waves from the
galactic disk to the solar neighborhood), then there is an ``experimental''
anticorrelation connection between the modulation of the ADM density in
the solar interior and the number of sunspots. Therefore, this is also true for
the relationship between the periods of S-star cycles and the sunspot cycles
(see ``experimental'' anticorrelated data in Fig.~25 
in~\cite{RusovDarkUniverse2021}).

%
%This means that the conditions of ADM (or more precisely, the asymmetric Higgs
%CDM) can lead to the absence of self-annihilation, which allows the
%accumulation of a large amount of ADM at the center of the Sun due to their
%capture by its gravitational field~\cite{Gould1987}. When the weakly
%interacting DM particles absorb energy in the hottest, central part of the
%core, they then move to a colder, more peripheral region, where they accumulate
%additional energy again and again before 
%scattering~\cite{GouldRaffelt1990a,GouldRaffelt1990b}. This reduces the
%temperature contrast in the central region and reduces the central temperature
%by several percent (see 
%e.g.~\cite{LopesSilk2002,LopesSilk2010,LopesSilk2012,Lopes2014}). Despite the
%small population of DM (at most $\sim 1$ particle per $10^{10}$ baryons), these
%small thermal gradient adjustments can have measurable effects on our own Sun
%(see~\cite{Vincent2016}).
%

This means that the conditions of ADM can lead to the absence of 
self-annihilation, which allows the
accumulation of a large amount of ADM at the center of the Sun due to their
capture by its gravitational field~\cite{Gould1987}. When the weakly
interacting DM particles absorb energy in the hottest, central part of the
core, they then move to a colder, more peripheral region, where they accumulate
additional energy again and again before 
scattering~\cite{GouldRaffelt1990a,GouldRaffelt1990b}. This reduces the
temperature contrast in the central region and reduces the central temperature
by several percent (see 
e.g.~\cite{LopesSilk2002,LopesSilk2010,LopesSilk2012,Lopes2014}). Despite the
small population of DM (at most $\sim 1$ particle per $10^{10}$ baryons), these
small thermal gradient adjustments can have measurable effects on our own Sun
(see~\cite{Vincent2016}).

%These include the solar structure itself -- (1) sunspots variations and 
%(2) deviations of sound speed 
%$\delta c_s / c_s = (c_{s,obs} - c_{s,th}) / c_{s,obs}$, both of which are
%related to solving the solar axion problem and density modulation of the Higgs
%CDM gravitationally captured in the depths of the Sun. At the same time, they
%are an ``experimental'' key to the the solution of the problem of the solar
%corona heating, associated with dark matter axions on the Sun and the asymmetry
%of the Higgs CDM halo and anticorrelated 11-year S -stars around the black hole
%(through the vertical density waves from the disk to the solar neighborhood;
%(see Eqs.~(27)-(28) and Fig.~25 in~\cite{RusovDarkUniverse2021}).

These include the solar structure itself -- (1) sunspots variations and 
(2) deviations of sound speed 
$\delta c_s / c_s = (c_{s,obs} - c_{s,th}) / c_{s,obs}$, both of which are
related to solving the solar axion problem and density modulation of the ADM
gravitationally captured in the depths of the Sun. At the same time, they
are an ``experimental'' key to the the solution of the problem of the solar
corona heating, associated with dark matter axions on the Sun and the ADM particles,
and anticorrelated 11-year S-stars around the black hole
(through the vertical density waves from the disk to the solar neighborhood;
(see Eqs.~(27)-(28) and Fig.~25 in~\cite{RusovDarkUniverse2021}).

%
%Let us first discuss the experimental data on the variability of sunspots. It
%is known that, according to Appendix~C in~\cite{RusovDarkUniverse2021}, the
%general laws of the theory of almost empty anchored magnetic flux tubes (MFT)
%with $B \sim 10^7 ~G$ have been developed, starting from the tachocline to the
%surface of the Sun. The main result of this theory is the formation of solar 
%axions and a magnetic O-loop inside the MFT near the tachocline. In this 
%magnetic O-shaped loop (based on the turbulent Kolmogorov cascade), axions are
%converted into photons, producing photons of axionic origin near the bottom of
%the convective zone, i.e. tachocline. On the other hand, high-energy photons 
%from the radiation zone through axion-photon oscillations in the O-loop inside
%the MFT near the tachocline create the so-called axions of photonic origin
%under the sunspot. This means that in such strong magnetic fields, the 
%Parker-Birman cooling effect of the MFT develops due to the ``disappearance''
%of the Parker convective heat transfer, and consequently, the temperature in
%the lower part of the magnetic tube with the help of axions of photon origin
%from photon-axion oscillations in O-loop near tachocline. As a result, a free 
%path opens up for photons of axion origin from the tachocline to the 
%photosphere! The latter means that the existence of photons of axion origin
%leads to the manifestation of identity between magnetic flux tubes and 
%sunspots!
%

Let us first discuss the experimental data on the variability of sunspots. It
is known that, according to Appendix~C in~\cite{RusovDarkUniverse2021}, the
general laws of the theory of almost empty anchored magnetic flux tubes (MFT)
with $B \sim 10^7 ~G$ have been developed, starting from the tachocline to the
surface of the Sun. The main result of this theory is the formation of solar 
axions and a magnetic O-loop inside the MFT near the tachocline. In this 
magnetic O-shaped loop (based on the turbulent Kolmogorov cascade), axions are
converted into photons, producing photons of axionic origin near the bottom of
the convective zone, i.e. tachocline. On the other hand, high-energy photons 
from the radiation zone through axion-photon oscillations in the O-loop inside
the MFT near the tachocline create the so-called axions of photonic origin
under the sunspot. This means that in such strong magnetic fields, the 
Parker-Birman cooling effect of the MFT develops due to the ``disappearance''
of the Parker convective heat transfer, and consequently, the temperature in
the lower part of the magnetic tube with the help of axions of photon origin
from photon-axion oscillations in O-loop near tachocline. As a result, a free 
path opens up for photons of axion origin from the tachocline to the 
photosphere! The latter means that the existence of photons of axion origin
leads to the manifestation of identity between magnetic flux tubes and 
sunspots!

%Since we know that the difference between the axion luminosity fractions 
%($L_a / L_{Sun} \so, 3.6 \cdot 10^{-4}$) and the 11-year Higgs CDM luminosity
%fractions (maximum for $(L_{HiggsCDM})_{+} / L_{Sun} \sim 3.24 \cdot 10^{-4}$
%(see $\sim 10^{-30}~erg/s$ in the right panel of the Fig.~4 
%in~\cite{Vincent2016}) and mimimum for 
%$(L_{HiggsCDM})_{-} / L_{Sun} \sim 3.24 \cdot 10^{-5}$ 
%(see $\sim 10^{-29}~erg/s$ in~\cite{RusovDarkUniverse2021}), as well as the
%probability $P_\gamma$, which describes the relative fraction of photons of 
%axion origin propagating along the magnetic tube, and consequently, the
%relative fraction of the 
%$\langle sunspot~area \rangle _{max} \approx 7.5 \cdot 10^9 ~km^2$ and
%$\langle sunspot~area \rangle _{min} \approx 9 \cdot 10^8 ~km^2$~\cite{Dikpati2008,Gough2010}
%for cycle 22, experimentally observed by the Japanese X-ray telescope 
%Yohkoh~(1991)~\cite{Zioutas2009}, predetermines theoretical equations 
%(Eqs.~(27)-(28) in~\cite{RusovDarkUniverse2021}) that allow to explain the
%maximum and minimum of the coronal luminosity fraction

\begin{figure}
  \begin{center}
    \includegraphics[width=8cm]{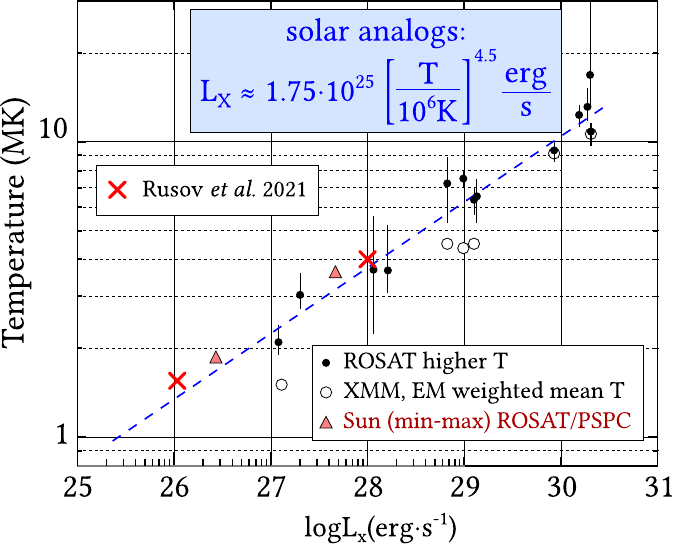}
  \end{center}
\caption{Dependence of the coronal temperature on the X-ray luminosity of solar
analogs (ROSAT, XMM-Newton, adopted from~\cite{Gudel2004}). ROSAT/PSPC and 
ASCA/SIS data (see Table~2 in~\cite{Peres2000}) are shown in comparison with
the maximum and minimum corona luminosity (see
%Fig.~\ref{fig-helioseismology} and
Fig.~2d in~\cite{RusovDarkUniverse2021}).}
\label{fig-luminosity-temperature}
\end{figure}

Since we know that the difference between the axion luminosity fractions 
($L_a / L_{Sun} \sim 3.6 \cdot 10^{-4}$) and the 11-year ADM luminosity
fractions (maximum for $(L_{ADM})_{+} / L_{Sun} \sim 3.24 \cdot 10^{-4}$
(see $\sim 10^{-30}~erg/s$ in the right panel of the Fig.~4 
in~\cite{Vincent2016}) and mimimum for 
$(L_{ADM})_{-} / L_{Sun} \sim 3.24 \cdot 10^{-5}$ 
(see $\sim 10^{-29}~erg/s$ in~\cite{RusovDarkUniverse2021}), as well as the
probability $P_\gamma$, which describes the relative fraction of photons of 
axion origin propagating along the magnetic tube, and consequently, the
relative fraction of the 
$\langle sunspot~area \rangle _{max} \approx 7.5 \cdot 10^9 ~km^2$ and
$\langle sunspot~area \rangle _{min} \approx 9 \cdot 10^8 ~km^2$~\cite{Dikpati2008,Gough2010}
for cycle 22, experimentally observed by the Japanese X-ray telescope 
Yohkoh~(1991)~\cite{Zioutas2009}, predetermines theoretical equations 
(Eqs.~(27)-(28) in~\cite{RusovDarkUniverse2021}) that allow to explain the
maximum and minimum of the coronal luminosity fraction
%
%
%\begin{equation}
%\frac{(L_{corona}^X)_{max}}{L_{Sun}} = \frac{(L_a^{*})_{max}}{L_{Sun}} \times
%(P_{\gamma})_{max} = \frac{L_a - (L_{HiggsCDM})_{-}}{L_{Sun}} \times 
%(P_{\gamma})_{max} \sim 2.4 \cdot 10^{-6},
%\label{eq7.5-sep-18}
%\end{equation}
%
%\begin{equation}
%\frac{(L_{corona}^X)_{min}}{L_{Sun}} = \frac{(L_a^{*})_{min}}{L_{Sun}} \times
%(P_{\gamma})_{min} = \frac{L_a - (L_{HiggsCDM})_{+}}{L_{Sun}} \times 
%(P_{\gamma})_{min} \sim 3.1 \cdot 10^{-8},
%\label{eq7.5-sep-19}
%\end{equation}
%

\begin{equation}
\frac{(L_{corona}^X)_{max}}{L_{Sun}} = \frac{(L_a^{*})_{max}}{L_{Sun}} \times
(P_{\gamma})_{max} = \frac{L_a - (L_{ADM})_{-}}{L_{Sun}} \times 
(P_{\gamma})_{max} \sim 2.4 \cdot 10^{-6},
\label{eq-PLB7-12}
%\label{eq7.5-sep-18}
\end{equation}

\begin{equation}
\frac{(L_{corona}^X)_{min}}{L_{Sun}} = \frac{(L_a^{*})_{min}}{L_{Sun}} \times
(P_{\gamma})_{min} = \frac{L_a - (L_{ADM})_{+}}{L_{Sun}} \times 
(P_{\gamma})_{min} \sim 3.1 \cdot 10^{-8},
\label{eq-PLB7-13}
%\label{eq7.5-sep-19}
\end{equation}

\noindent which are good enough. Still, our estimates of the theoretical
variations of solar luminosity 
$(L_{corona}^X)_{max} \sim 1.0 \cdot 10^{28}~erg/s$ at a
temperature $T_{max} \sim 4\cdot 10^6~K$ and 
$(L_{corona}^X)_{min} \sim 1.2 \cdot 10^{26}~erg/s$ at a temperature
$T_{min} \sim 1.5\cdot 10^6~K$ are somewhat better than the data 
in~\cite{Peres2000} (see our Fig.~\ref{fig-luminosity-temperature}, and
Fig.~2d in~\cite{RusovDarkUniverse2021}).

%And finally, remembering the amazing words of Robert Dicke, we understand that
%the 11-year variation of the Higgs CDM luminosity fraction on the Sun is not
%only the cause of the 11-year variation of the luminosity of S-stars around
%a black hole, but also a consequence of the 11-year variation of the solar
%axions fraction in the luminosity. The latter (axions) are not only the cause
%of the identity of both magnetic flux tubes and sunspots, but also a remarkable
%cause of the identity between solar axions and warm dark matter axion remnants
%in the galactic Higgs CDM halo (see Eq.~(\ref{eq-5.2-09})).

And finally, remembering the amazing words of Robert Dicke, we understand that
the 11-year variation of the ADM luminosity fraction on the Sun is not
only the cause of the 11-year variation of the luminosity of S-stars around
a black hole, but also a consequence of the 11-year variation of the solar
axions fraction in the luminosity. The latter (axions) are not only the cause
of the identity of both magnetic flux tubes and sunspots, but also a remarkable
cause of the identity between solar axions and coherent axions of warm dark
in the galactic ADM halo (see Eq.~(\ref{eq-PLB7-09})).

%
%The second example is related to the experimental data on the deviation of
%sound speed $\delta c_s / c_s = (c_{s,obs} - c_{s,th}) / c_{s,obs}$ and 
%represents a solution to the problem of the solar axion and the density
%modulation of the Higgs CDM gravitationally captured in the depths of the Sun.
%

The second example is related to the experimental data on the deviation of
sound speed $\delta c_s / c_s = (c_{s,obs} - c_{s,th}) / c_{s,obs}$ and 
represents a solution to the problem of the solar axion and the 11-year density
modulation of the ADM gravitationally captured in the depths of the Sun.

In order to determine the values of the 11-year variations in the local density
of ADM on the Sun, we use the results of Aaron C. Vincent~\cite{Vincent2020},
who shows (see right panel in Fig.~2 in~\cite{Vincent2020}) a series of 
trajectories for the 1M star interacting with DM with densities varying
from $1 - 10^4$ times the local DM density $\rho_0 = 0.38 ~GeV cm^{-3}$,
produced using the MESA~\cite{Paxton2010} stellar evolution software and the
Spergel and Press approach~\cite{Spergel1985}.

In contrast to the constant local of DM density of 
$\rho_0 = 0.38 ~GeV cm^{-3}$, we proposed the value of 11-year variations in
the local ADM density on the Sun, associated with a kink in the 
Luminosity-Temperature curve ($L(L_{Sun}) - T(K)$; see right panel in Fig.~2 
in~\cite{Vincent2020}), in the form of a local density minimum 
$\rho _{ADM} ^{min} = \rho_0 / 3.19 \approx 0.1 ~GeV/cm^3$, and a maximum
$\rho _{ADM} ^{max} = \rho_0 \times 3.19 \approx 1.0 ~GeV/cm^3$.

%We noticed that the existence of a solar maximum (with the maximum fraction of
%the axion luminosity $(L_a^*)_{max} \gg (L_{HiggsCDM})_{-}$ and a solar 
%minimum (with the minimum fraction of the axion luminosity 
%$(L_a^*)_{min} \ll (L_{HiggsCDM})_{+}$ leads to to the existence of an 11-year
%variation in the uncertainty of the deviation of radial sound speed profile, 
%for which the maximum and minimum neutrino fluxes can simultaneously solve
%a very complex ``problem of solar composition'', or otherwise called ``the
%problem of solar abundance'', which is associated with the solar cycle.
%Most important here is that the neutrino flux from the subdominant CNO cycle is
%linearly dependent on the metallicity ($Z$) of the solar core (see 
%e.g.~\cite{Gann2015,Borexino2020a,Borexino2020b,TapiaArellano2021}). 
%It means that the modulation of solar abundance, with high metallicity at the
%solar maximum ($Z\sim 0.0170$ with $Z/X=0.023$~\citep{Grevesse1998}) and
%low metallicity at the solar minimum ($Z\sim 0.0133$ with 
%$Z/X=0.0178$~\citep{Asplund2009}), is in good agreement with the
%deviation of the radial sound speed profile 
%$\delta c_s / c_s = (c_{s,obs} - c_{s,th})/c_{s,obs}$ inside the Sun 
%(Fig.~\ref{fig-helioseismology}), which consists of two related models:
%for the maximum luminosity of axions (Fig.~\ref{fig-helioseismology}a)
%and for the minimum luminosity of axions, which is associated with the maximum
%luminosity of Higgs CDM (Fig.~\ref{fig-helioseismology}b).

\begin{figure}[tb!]
%\begin{figure}
\begin{center}
  \includegraphics[width=16cm]{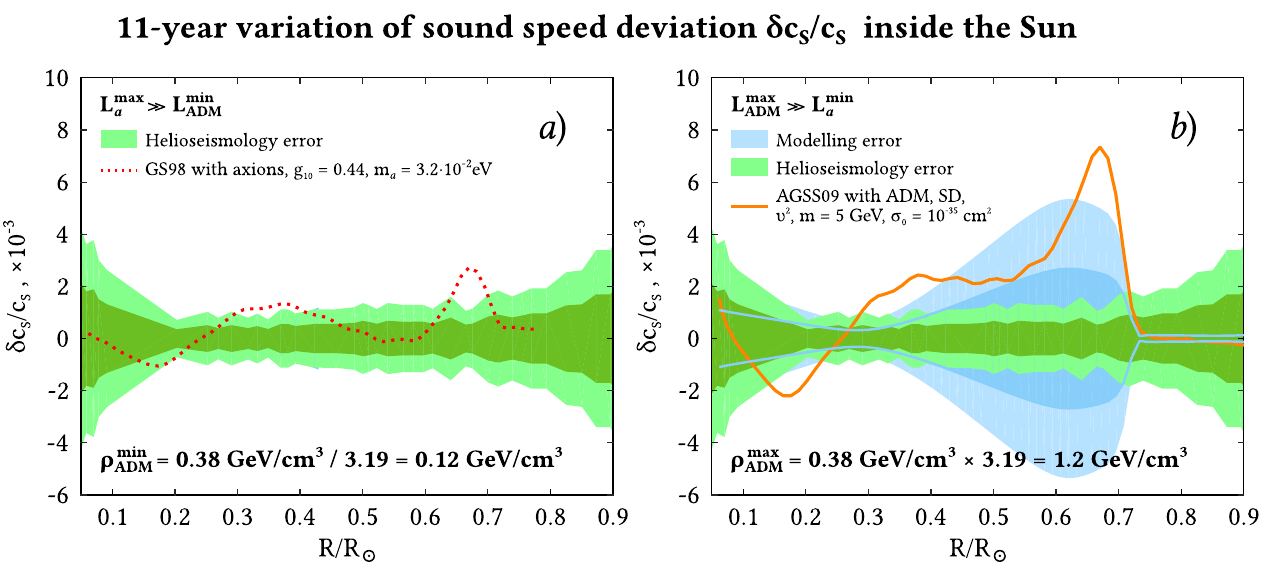}
\end{center}
\caption{Deviation of the radial speed of sound profile (Sun - model)/Sun in
the solar interior from values obtained for two models: (a) GS98  with axions
(the red dots from the values of \cite{Grevesse1998}) almost coincide with the values 
in~\cite{Vinyoles2015}
and (b) the ADM captured inside the Sun (solid orange
line of ADM~\citep{Vincent2016}).
Colored areas indicate 1$\sigma$ and 2$\sigma$ errors in modeling (thick blue
bar;~\cite{Vincent2015a,Vincent2015b}) and helioseismological inverses (thinner
green bar;~\cite{DeglInnoccenti1997,Fiorentini2001}).
The deviation of the radial speed of sound is associated with the part of the
corona luminosity~\citep{RusovDarkUniverse2021} in the total luminosity of the
Sun, which can be easily determined from (a) the solar maximum (involving the
maximum axion luminosity fraction ($(L_a^*)_{max} \gg (L_{ADM})_{-}$) and
(b) the solar minimum (involving minimum fraction of the axion luminosity 
($(L_a^*)_{min} \ll (L_{ADM})_{+}$) and is in good agreement with
nonstandard solar model.}
\label{fig-helioseismology}
\end{figure}

We noticed that the existence of a solar maximum (with the maximum fraction of
the axion luminosity $(L_a^*)_{max} \gg (L_{ADM})_{-}$ and a solar 
minimum (with the minimum fraction of the axion luminosity 
$(L_a^*)_{min} \ll (L_{ADM})_{+}$ leads to to the existence of an 11-year
variation in the uncertainty of the deviation of radial sound speed profile, 
for which the maximum and minimum neutrino fluxes can simultaneously solve
a very complex ``problem of solar composition'', or otherwise called ``the
problem of solar abundance'', which is associated with the solar cycle.
Most important here is that the neutrino flux from the subdominant CNO cycle is
linearly dependent on the metallicity ($Z$) of the solar core (see 
e.g.~\cite{Gann2015,Borexino2020a,Borexino2020b,TapiaArellano2021}). 
It means that the modulation of solar abundance, with high metallicity at the
solar maximum ($Z\sim 0.0170$ with $Z/X=0.023$~\citep{Grevesse1998}) and
low metallicity at the solar minimum ($Z\sim 0.0133$ with 
$Z/X=0.0178$~\citep{Asplund2009}), is in good agreement with the
deviation of the radial sound speed profile 
$\delta c_s / c_s = (c_{s,obs} - c_{s,th})/c_{s,obs}$ inside the Sun 
(Fig.~\ref{fig-helioseismology}), which consists of two related models:
for the maximum luminosity of axions (Fig.~\ref{fig-helioseismology}a)
and for the minimum luminosity of axions, which is associated with the maximum
luminosity of ADM (Fig.~\ref{fig-helioseismology}b).

%
%At the same time, we understand that, regarding the 11-year variation in the
%deviation of the radial sound speed profile $\delta c_s / c_s$ inside the Sun, 
%we did not make our own calculations, but adopted the ready-made calculations
%in the form of two interlinked 5.5-year models: the maximum luminosity of
%axions $(L_a^*)_{max}$ (Fig.~\ref{fig-helioseismology}a; see 
%Fig.~5 in~\cite{Vinyoles2015}) and the maximum luminosity of the Higgs CDM 
%$(L_{HiggsCDM})_{+}$ (Fig.~\ref{fig-helioseismology}b; see Fig.~6 
%in~\cite{Vincent2016}). It can be seen from the fits in 
%Fig.~\ref{fig-helioseismology} that the resulting non-standard solar model is
%good, but not the best, and can become better 
%%\hl{when the problem of solar
%%abundance, associated with taking into account the measured solar (high and
%%low) metallicity, is apparently solved because of the fundamental physics of
%%the Sun} \hl{[Непонятно, что мы тут хотим сказать]}.
%when the problem of solar abundance, related to the measured solar variations
%(high and low) metallicity, linearly depending on the neutrino flux of the
%subdominant CNO cycle, is a complete solution of fundamental physics of the Sun.

At the same time, we understand that, regarding the 11-year variation in the
deviation of the radial sound speed profile $\delta c_s / c_s$ inside the Sun, 
we did not make our own calculations, but adopted the ready-made calculations
in the form of two interlinked 5.5-year models: the maximum luminosity of
axions $L_a^{max}$ (Fig.~\ref{fig-helioseismology}a; see 
Fig.~5 in~\cite{Vinyoles2015}) and the maximum luminosity of the ADM 
$L_{ADM}^{max}$ (Fig.~\ref{fig-helioseismology}b; see Fig.~6 
in~\cite{Vincent2016}). It can be seen from the fits in 
Fig.~\ref{fig-helioseismology} that the resulting non-standard solar model is
good, but not the best, and can become better 
when the problem of solar abundance, related to the measured solar variations
of (high and low) metallicity, and surprisingly, anticorrelated to the 11-year
variations of local ADM density, linearly depending on the neutrino flux of the
subdominant CNO cycle, is a complete solution of fundamental physics of the Sun.

%
%In addition to all other, the resulting nonstandard solar model is also a
%remarkable explanation of the reason for the identity between solar axions and
%warm dark matter axion remnants in the galactic Higgs CDM halo (see 
%Eq.~(\ref{eq-5.2-09})).
%
From here, we understand that the experimental data of the non-standard solar
model is an excellent explanation of the reason for the identity between solar axions and
warm dark matter axion remnants in the galactic ADM halo (see 
Section~\ref{sec-axions-identity-proof}), as well as for the formation of ADM
in the Universe caused by the decay of Higgs dark matter (see Section~\ref{sec-ADM-Higgs-proof})!

\section{Conclusions}
\label{sec-conclusions}
%
%Based on the modified Turner model~\cite{Turner1987,Turner1986Report} in the early Universe,
%we have obtained a convincing evidence of the identity of asymmetric Higgs cold dark matter
%halo near a black hole, and the asymmetric dark matter (ADM) in
%the solar neighborhood~\cite{RusovDarkUniverse2021}, as well as the remarkable
%identity of solar axions (see Fig.~\ref{fig-helioseismology}) and the coherent
%axions of warm dark matter in the Universe (see 
%$m_a \sim 3.2 \cdot 10^{-2}~eV$).
%At the same time, it must be remembered that the initial axions with a mass of
%40~eV, which are the low-energy residue of the Peccei-Quinn solution to the strong
%CP problem~\cite{PecceiQuinn1977,PecceiQuinn1977PRD,Wilczek1978,Weinberg1978},
%make the main contribution to the energy density of the Universe (see 
%$\Omega_{axionHDM} \sim 0.31$ in Fig.~\ref{fig-axion-inflation} and 
%Eq.~(\ref{eq-5.1-R4-02})). Although it is not accessible for direct
%cosmological searches for hot dark matter axions with mass of 40~eV, the other one ($\Omega_{axionWDM}^{coherent} \sim 4.5 \cdot 10^{-5}$) is accessible for direct astrophysical experiments for axions of warm dark matter with a mass of 
%$m_a \sim 3.2 \cdot 10^{-2}~eV$, identical to solar axions~\cite{RusovDarkUniverse2021}!

Based on the modified Turner model~\cite{Turner1987,Turner1986Report} in the
early Universe, we have obtained a convincing evidence of both the formation of
ADM in the Universe (see Fig.~\ref{fig-helioseismology}b) caused by the decay
of Higgs dark matter without the so-called ``fifth'' carrier force for the 
Higgs field in the  Universe, and the remarkable identity of solar axions (see 
Fig.~\ref{fig-helioseismology}a) and the coherent axions of warm dark matter in
the Universe (see $m_a \sim 3.2 \cdot 10^{-2}~eV$ and 
Fig.~\ref{fig-axion-constraints}).
At the same time, it must be remembered that the initial thermal axions of hot
dark matter with a mass of $\sim 6.0~eV$, which are the low-energy residue of
the Peccei-Quinn solution to the strong CP 
problem~\cite{PecceiQuinn1977,PecceiQuinn1977PRD,Wilczek1978,Weinberg1978},
make the main contribution to the energy density of the Universe (see 
$\Omega_{axionHDM} \sim 0.32$ in Fig.~\ref{fig-axion-inflation} and 
Eq.~(\ref{eq-PLB7-06})). Although thermal axions of hot dark matter with a mass
of $\sim 6.0~eV$ are not accessible for direct
cosmological searches,
%for hot dark matter axions with mass of $\sim 6.0~eV$,
%but it is accessible for direct cosmological searches,
but direct astrophysical experiments are possible for axions of warm
dark matter with a mass of $m_a \sim 3.2 \cdot 10^{-2}~eV$ 
($\Omega_{axionWDM}^{coherent} \sim 4.5 \cdot 10^{-5}$), identical to solar
axions (see Fig.~\ref{fig-axion-constraints} and~\cite{RusovDarkUniverse2021}),
and the ADM particles with a mass $\sim 5~GeV$ in the Universe 
($\Omega_{ADM} \sim 0.27$), appearing from the decay of Higgs dark matter!


%\bibliographystyle{unsrtnat}
\bibliographystyle{elsarticle-num}
\bibliography{Rusov-AxionSunLuminosity}

\begin{thebibliography}{100}
\expandafter\ifx\csname url\endcsname\relax
  \def\url#1{\texttt{#1}}\fi
\expandafter\ifx\csname urlprefix\endcsname\relax\def\urlprefix{URL }\fi
\expandafter\ifx\csname href\endcsname\relax
  \def\href#1#2{#2} \def\path#1{#1}\fi

\bibitem{Turner1987}
M.~S. Turner, Early-universe thermal production of not-so-invisible axions,
  Phys. Rev. Lett. 59 (1987) 2489--2492, erratum Phys. Rev. Lett. 60, 1101
  (1988).
\newblock \href {https://doi.org/10.1103/PhysRevLett.59.2489}
  {\path{doi:10.1103/PhysRevLett.59.2489}}.

\bibitem{Turner1986Report}
M.~Turner, Thermal production of not so invisible axions in the early universe,
  Tech. Rep. FNAL/Pub-86/150-A, 6926041, Fermi National Accelerator Laboratory
  (October 1986).
\newblock \href {https://doi.org/10.2172/6926041} {\path{doi:10.2172/6926041}}.

\bibitem{RusovDarkUniverse2021}
V.~D. Rusov, I.~V. Sharph, V.~P. Smolyar, M.~V. Eingorn, M.~E. Beglaryan,
  Coronal heating problem solution by means of axion origin photons, Physics of
  the Dark Universe 31 (2021) 100746.
\newblock \href {https://doi.org/10.1016/j.dark.2020.100746}
  {\path{doi:10.1016/j.dark.2020.100746}}.

\bibitem{ref01}
G.~G. Raffelt, Axions and other very light bosons: {Part II} (astrophysical
  constraints), Phys. Lett. 592 (2004) 391.

\bibitem{ref02}
G.~G. Raffelt, Astrophysical axion bounds, Lect. Notes Phys. 741 (2008) 51--71,
  \href{http://arxiv.org/abs/hep-ph/0611350}{arXiv:hep-ph/0611350}.

\bibitem{PecceiQuinn1977}
R.~D. Peccei, H.~R. Quinn, Cp conservation in the presence of pseudoparticles,
  Physical Review Letters 38~(25) (1977) 1440–1443.
\newblock \href {https://doi.org/10.1103/PhysRevLett.38.1440}
  {\path{doi:10.1103/PhysRevLett.38.1440}}.

\bibitem{PecceiQuinn1977PRD}
R.~D. Peccei, H.~R. Quinn, Constraints imposed by $\mathrm{CP}$ conservation in
  the presence of pseudoparticles, Phys. Rev. D 16 (1977) 1791--1797.
\newblock \href {https://doi.org/10.1103/PhysRevD.16.1791}
  {\path{doi:10.1103/PhysRevD.16.1791}}.

\bibitem{Wilczek1978}
F.~Wilczek, Problem of strong $p$ and $t$ invariance in the presence of
  instantons, Phys. Rev. Lett. 40 (1978) 279--282.
\newblock \href {https://doi.org/10.1103/PhysRevLett.40.279}
  {\path{doi:10.1103/PhysRevLett.40.279}}.

\bibitem{Weinberg1978}
S.~Weinberg, A new light boson?, Physical Review Letters 40~(4) (1978)
  223–226.
\newblock \href {https://doi.org/10.1103/PhysRevLett.40.223}
  {\path{doi:10.1103/PhysRevLett.40.223}}.

\bibitem{Preskill1983}
J.~Preskill, M.~B. Wise, F.~Wilczek, Cosmology of the invisible axion, Physics
  Letters B 120~(1-3) (1983) 127--132.
\newblock \href {https://doi.org/10.1016/0370-2693(83)90637-8}
  {\path{doi:10.1016/0370-2693(83)90637-8}}.

\bibitem{Abbott1983}
L.~Abbott, P.~Sikivie, A cosmological bound on the invisible axion, Physics
  Letters B 120~(1-3) (1983) 133--136.
\newblock \href {https://doi.org/10.1016/0370-2693(83)90638-X}
  {\path{doi:10.1016/0370-2693(83)90638-X}}.

\bibitem{Dine1983}
M.~Dine, W.~Fischler, The not-so-harmless axion, Physics Letters B 120~(1-3)
  (1983) 137--141.
\newblock \href {https://doi.org/10.1016/0370-2693(83)90639-1}
  {\path{doi:10.1016/0370-2693(83)90639-1}}.

\bibitem{Kawasaki2013}
M.~Kawasaki, K.~Nakayama, Axions: Theory and cosmological role, Annual Review
  of Nuclear and Particle Science 63~(1) (2013) 69–95.
\newblock \href {https://doi.org/10.1146/annurev-nucl-102212-170536}
  {\path{doi:10.1146/annurev-nucl-102212-170536}}.

\bibitem{Marsh2016}
D.~J. Marsh, Axion cosmology, Physics Reports 643 (2016) 1 -- 79, axion
  cosmology.
\newblock \href
  {https://doi.org/http://dx.doi.org/10.1016/j.physrep.2016.06.005}
  {\path{doi:http://dx.doi.org/10.1016/j.physrep.2016.06.005}}.

\bibitem{DiLuzio2020}
L.~Di~Luzio, M.~Giannotti, E.~Nardi, L.~Visinelli, The landscape of qcd axion
  models, Physics Reports 870 (2020) 1–117.
\newblock \href {https://doi.org/10.1016/j.physrep.2020.06.002}
  {\path{doi:10.1016/j.physrep.2020.06.002}}.

\bibitem{Sikivie2021}
P.~Sikivie, Invisible axion search methods, Reviews of Modern Physics 93~(1)
  (2021) 015004.
\newblock \href {https://doi.org/10.1103/RevModPhys.93.015004}
  {\path{doi:10.1103/RevModPhys.93.015004}}.

\bibitem{Ayala2014}
A.~Ayala, I.~Dom\'{i}nguez, M.~Giannotti, A.~Mirizzi, O.~Straniero, Revisiting
  the bound on axion-photon coupling from globular clusters, Phys. Rev. Lett.
  113 (2014) 191302.
\newblock \href {https://doi.org/10.1103/PhysRevLett.113.191302}
  {\path{doi:10.1103/PhysRevLett.113.191302}}.

\bibitem{Buschmann2021}
M.~Buschmann, R.~T. Co, C.~Dessert, B.~R. Safdi, Axion emission can explain a
  new hard x-ray excess from nearby isolated neutron stars, Phys. Rev. Lett.
  126 (2021) 021102.
\newblock \href {https://doi.org/10.1103/PhysRevLett.126.021102}
  {\path{doi:10.1103/PhysRevLett.126.021102}}.

\bibitem{Carenza2019}
P.~Carenza, T.~Fischer, M.~Giannotti, G.~Guo, G.~Mart\'{i}nez-Pinedo,
  A.~Mirizzi, Improved axion emissivity from a supernova via nucleon-nucleon
  bremsstrahlung, Journal of Cosmology and Astroparticle Physics 2019~(10)
  (2019) 016–016.
\newblock \href {https://doi.org/10.1088/1475-7516/2019/10/016}
  {\path{doi:10.1088/1475-7516/2019/10/016}}.

\bibitem{Carenza2021}
P.~Carenza, B.~Fore, M.~Giannotti, A.~Mirizzi, S.~Reddy, Enhanced supernova
  axion emission and its implications, Phys. Rev. Lett. 126 (2021) 071102.
\newblock \href {https://doi.org/10.1103/PhysRevLett.126.071102}
  {\path{doi:10.1103/PhysRevLett.126.071102}}.

\bibitem{CAST2017}
{CAST Collaboration}, New cast limit on the axion–photon interaction, Nature
  Physics 13~(6) (2017) 584–590.
\newblock \href {https://doi.org/10.1038/nphys4109}
  {\path{doi:10.1038/nphys4109}}.

\bibitem{Asztalos2010}
S.~J. Asztalos, G.~Carosi, C.~Hagmann, D.~Kinion, , K.~van Bibber, J.~Hoskins,
  J.~Hwang, P.~Sikivie, D.~B. Tanner, R.~Bradley, J.~Clarke, {SQUID}-based
  microwave cavity search for dark-matter axions, Phys. Rev. Lett. 104 (2010)
  041301.

\bibitem{Wuensch1989}
W.~U. Wuensch, S.~D. Panfilis-Wuensch, Y.~K. Semertzidis, J.~T. Rogers, A.~C.
  Melissinos, H.~J. Halama, B.~E. Moskowitz, A.~G. Prodell, W.~B. Fowler, F.~A.
  Nezrick, Results of a laboratory search for cosmic axions and other weakly
  coupled light particles, Phys. Rev. D 40~(10) (1989) 3153--3167.

\bibitem{Brubaker2017}
B.~M. Brubaker, L.~Zhong, S.~K. Lamoreaux, K.~W. Lehnert, K.~A. van Bibber,
  {HAYSTAC} axion search analysis procedure, Physical Review D 96~(12) (2017)
  123008.
\newblock \href {https://doi.org/10.1103/PhysRevD.96.123008}
  {\path{doi:10.1103/PhysRevD.96.123008}}.

\bibitem{Zhong2018}
L.~Zhong, S.~Al~Kenany, K.~M. Backes, B.~M. Brubaker, S.~B. Cahn, G.~Carosi,
  Y.~V. Gurevich, W.~F. Kindel, S.~K. Lamoreaux, K.~W. Lehnert, S.~M. Lewis,
  M.~Malnou, R.~H. Maruyama, D.~A. Palken, N.~M. Rapidis, J.~R. Root,
  M.~Simanovskaia, T.~M. Shokair, D.~H. Speller, I.~Urdinaran, K.~A. van
  Bibber, Results from phase 1 of the haystac microwave cavity axion
  experiment, Phys. Rev. D 97 (2018) 092001.
\newblock \href {https://doi.org/10.1103/PhysRevD.97.092001}
  {\path{doi:10.1103/PhysRevD.97.092001}}.

\bibitem{Braine2020}
T.~Braine, R.~Cervantes, N.~Crisosto, N.~Du, S.~Kimes, L.~J. Rosenberg,
  G.~Rybka, J.~Yang, D.~Bowring, A.~S. Chou, R.~Khatiwada, A.~Sonnenschein,
  W.~Wester, G.~Carosi, N.~Woollett, L.~D. Duffy, R.~Bradley, C.~Boutan,
  M.~Jones, B.~H. LaRoque, N.~S. Oblath, M.~S. Taubman, J.~Clarke, A.~Dove,
  A.~Eddins, S.~R. O'Kelley, S.~Nawaz, I.~Siddiqi, N.~Stevenson, A.~Agrawal,
  A.~V. Dixit, J.~R. Gleason, S.~Jois, P.~Sikivie, J.~A. Solomon, N.~S.
  Sullivan, D.~B. Tanner, E.~Lentz, E.~J. Daw, J.~H. Buckley, P.~M. Harrington,
  E.~A. Henriksen, K.~W. Murch, Extended search for the invisible axion with
  the axion dark matter experiment, Phys. Rev. Lett. 124 (2020) 101303.
\newblock \href {https://doi.org/10.1103/PhysRevLett.124.101303}
  {\path{doi:10.1103/PhysRevLett.124.101303}}.

\bibitem{Lee2020}
S.~Lee, S.~Ahn, J.~Choi, B.~R. Ko, Y.~K. Semertzidis, Axion dark matter search
  around $6.7\mu${eV}, Phys. Rev. Lett. 124 (2020) 101802.
\newblock \href {https://doi.org/10.1103/PhysRevLett.124.101802}
  {\path{doi:10.1103/PhysRevLett.124.101802}}.

\bibitem{Backes2021}
K.~M. Backes, D.~A. Palken, S.~A. Kenany, B.~M. Brubaker, S.~B. Cahn,
  A.~Droster, G.~C. Hilton, S.~Ghosh, H.~Jackson, S.~K. Lamoreaux, et~al., A
  quantum enhanced search for dark matter axions, Nature 590~(7845) (2021)
  238--242.
\newblock \href {https://doi.org/10.1038/s41586-021-03226-7}
  {\path{doi:10.1038/s41586-021-03226-7}}.

\bibitem{Raffelt2008}
G.~G. Raffelt, Astrophysical Axion Bounds, Springer Berlin Heidelberg, Berlin,
  Heidelberg, 2008, pp. 51--71.
\newblock \href {https://doi.org/10.1007/978-3-540-73518-2_3}
  {\path{doi:10.1007/978-3-540-73518-2_3}}.

\bibitem{Carosi2013}
G.~Carosi, A.~Friedland, M.~Giannotti, M.~Pivovarov, J.~Ruz, J.~Vogel, Probing
  the axion-photon coupling: phenomenological and experimental perspectives. a
  snowmass white paper, \href{http://arxiv.org/abs/1309.7035}{arXiv:1309.7035}
  (2013).

\bibitem{Kim1979}
J.~E. Kim, Weak-interaction singlet and strong $\mathrm{CP}$ invariance, Phys.
  Rev. Lett. 43 (1979) 103--107.
\newblock \href {https://doi.org/10.1103/PhysRevLett.43.103}
  {\path{doi:10.1103/PhysRevLett.43.103}}.

\bibitem{Shifman1980}
M.~Shifman, A.~Vainshtein, V.~Zakharov, Can confinement ensure natural {CP}
  invariance of strong interactions?, Nucl. Phys. B 166 (1980) 493.

\bibitem{Friedland2013}
A.~Friedland, M.~Giannotti, M.~Wise, Constraining the axion-photon coupling
  with massive stars, Phys. Rev. Lett. 110 (2013) 061101.
\newblock \href {https://doi.org/10.1103/PhysRevLett.110.061101}
  {\path{doi:10.1103/PhysRevLett.110.061101}}.

\bibitem{Graham2015}
P.~W. Graham, I.~G. Irastorza, S.~K. Lamoreaux, A.~Lindner, K.~A. van Bibber,
  Experimental searches for the axion and axion-like particles, Annual Review
  of Nuclear and Particle Science 65~(1) (2015) 485–514.
\newblock \href {https://doi.org/10.1146/annurev-nucl-102014-022120}
  {\path{doi:10.1146/annurev-nucl-102014-022120}}.

\bibitem{Moroi1998}
T.~Moroi, H.~Murayama, Axionic hot dark matter in the hadronic axion window,
  Physics Letters B 440~(1) (1998) 69--76.
\newblock \href {https://doi.org/10.1016/S0370-2693(98)01091-0}
  {\path{doi:10.1016/S0370-2693(98)01091-0}}.

\bibitem{PlanckCol2018}
{Planck Collaboration}, Planck 2018 results - {VI}. cosmological parameters,
  A\&A 641 (2020) A6.
\newblock \href {https://doi.org/10.1051/0004-6361/201833910}
  {\path{doi:10.1051/0004-6361/201833910}}.

\bibitem{Aiola2020}
S.~Aiola, E.~Calabrese, L.~Maurin, S.~Naess, B.~L. Schmitt, M.~H. Abitbol,
  G.~E. Addison, P.~A.~R. Ade, D.~Alonso, M.~Amiri, S.~Amodeo, E.~Angile, J.~E.
  Austermann, T.~Baildon, N.~Battaglia, J.~A. Beall, R.~Bean, D.~T. Becker,
  J.~R. Bond, S.~M. Bruno, V.~Calafut, L.~E. Campusano, F.~Carrero, G.~E.
  Chesmore, H.~mei Cho, S.~K. Choi, S.~E. Clark, N.~F. Cothard, D.~Crichton,
  K.~T. Crowley, O.~Darwish, R.~Datta, E.~V. Denison, M.~J. Devlin, C.~J.
  Duell, S.~M. Duff, A.~J. Duivenvoorden, J.~Dunkley, R.~Dünner,
  T.~Essinger-Hileman, M.~Fankhanel, S.~Ferraro, A.~E. Fox, B.~Fuzia, P.~A.
  Gallardo, V.~Gluscevic, J.~E. Golec, E.~Grace, M.~Gralla, Y.~Guan, K.~Hall,
  M.~Halpern, D.~Han, P.~Hargrave, M.~Hasselfield, J.~M. Helton, S.~Henderson,
  B.~Hensley, J.~C. Hill, G.~C. Hilton, M.~Hilton, A.~D. Hincks, R.~Hložek,
  S.-P.~P. Ho, J.~Hubmayr, K.~M. Huffenberger, J.~P. Hughes, L.~Infante,
  K.~Irwin, R.~Jackson, J.~Klein, K.~Knowles, B.~Koopman, A.~Kosowsky,
  V.~Lakey, D.~Li, Y.~Li, Z.~Li, M.~Lokken, T.~Louis, M.~Lungu, A.~MacInnis,
  M.~Madhavacheril, F.~Maldonado, M.~Mallaby-Kay, D.~Marsden, J.~McMahon,
  F.~Menanteau, K.~Moodley, T.~Morton, T.~Namikawa, F.~Nati, L.~Newburgh, J.~P.
  Nibarger, A.~Nicola, M.~D. Niemack, M.~R. Nolta, J.~Orlowski-Sherer, L.~A.
  Page, C.~G. Pappas, B.~Partridge, P.~Phakathi, G.~Pisano, H.~Prince,
  R.~Puddu, F.~J. Qu, J.~Rivera, N.~Robertson, F.~Rojas, M.~Salatino,
  E.~Schaan, A.~Schillaci, N.~Sehgal, B.~D. Sherwin, C.~Sierra, J.~Sievers,
  C.~Sifon, P.~Sikhosana, S.~Simon, D.~N. Spergel, S.~T. Staggs, J.~Stevens,
  E.~Storer, D.~D. Sunder, E.~R. Switzer, B.~Thorne, R.~Thornton, H.~Trac,
  J.~Treu, C.~Tucker, L.~R. Vale, A.~V. Engelen, J.~V. Lanen, E.~M. Vavagiakis,
  K.~Wagoner, Y.~Wang, J.~T. Ward, E.~J. Wollack, Z.~Xu, F.~Zago, N.~Zhu, The
  {Atacama} cosmology telescope: {DR4} maps and cosmological parameters,
  Journal of Cosmology and Astroparticle Physics 2020~(12) (2020) 047.
\newblock \href {https://doi.org/10.1088/1475-7516/2020/12/047}
  {\path{doi:10.1088/1475-7516/2020/12/047}}.

\bibitem{PDG2020}
{Particle Data Group}, {Review of Particle Physics}, Progress of Theoretical
  and Experimental Physics 2020~(8), 083C01 (08 2020).
\newblock \href {https://doi.org/10.1093/ptep/ptaa104}
  {\path{doi:10.1093/ptep/ptaa104}}.

\bibitem{PlanckCol2015}
{Planck Collaboration}, Planck 2015 results - {XIII}. cosmological parameters,
  A\&A 594 (2016) A13.
\newblock \href {https://doi.org/10.1051/0004-6361/201525830}
  {\path{doi:10.1051/0004-6361/201525830}}.

\bibitem{Patrignani2016}
C.~{Patrignani}, {Particle Data Group}, {Review of Particle Physics}, Chinese
  Physics C 40~(10) (2016) 100001, 11. STATUS OF HIGGS BOSON PHYSICS.
\newblock \href {https://doi.org/10.1088/1674-1137/40/10/100001}
  {\path{doi:10.1088/1674-1137/40/10/100001}}.

\bibitem{DeSimone2019}
A.~De~Simone, Introduction to cosmology and dark matter, in: Proceedings of the
  2018 European School of High-Energy Physics, Maratea, Italy, 2019.
\newblock \href {https://doi.org/10.23730/CYRSP-2019-006.145}
  {\path{doi:10.23730/CYRSP-2019-006.145}}.

\bibitem{Giddings1988}
S.~B. Giddings, A.~Strominger, Loss of incoherence and determination of
  coupling constants in quantum gravity, Nuclear Physics B 307~(4) (1988)
  854--866.
\newblock \href {https://doi.org/10.1016/0550-3213(88)90109-5}
  {\path{doi:10.1016/0550-3213(88)90109-5}}.

\bibitem{Coleman1988}
S.~Coleman, Why there is nothing rather than something: A theory of the
  cosmological constant, Nuclear Physics B 310~(3) (1988) 643--668.
\newblock \href {https://doi.org/10.1016/0550-3213(88)90097-1}
  {\path{doi:10.1016/0550-3213(88)90097-1}}.

\bibitem{Gilbert1989}
G.~Gilbert, Wormhole-induced proton decay, Nuclear Physics B 328~(1) (1989)
  159--170.
\newblock \href {https://doi.org/10.1016/0550-3213(89)90097-7}
  {\path{doi:10.1016/0550-3213(89)90097-7}}.

\bibitem{Harlow2019}
D.~Harlow, H.~Ooguri, Constraints on symmetries from holography, Phys. Rev.
  Lett. 122 (2019) 191601.
\newblock \href {https://doi.org/10.1103/PhysRevLett.122.191601}
  {\path{doi:10.1103/PhysRevLett.122.191601}}.

\bibitem{Harlow2019arxiv}
D.~Harlow, H.~Ooguri, Symmetries in quantum field theory and quantum gravity
  (2019).
\newblock \href {http://arxiv.org/abs/1810.05338} {\path{arXiv:1810.05338}}.

\bibitem{Co2020}
R.~T. Co, K.~Harigaya, Axiogenesis, Phys. Rev. Lett. 124 (2020) 111602.
\newblock \href {https://doi.org/10.1103/PhysRevLett.124.111602}
  {\path{doi:10.1103/PhysRevLett.124.111602}}.

\bibitem{Harigaya2020Conf}
K.~Harigaya, Axion kinetic misalignment and baryogenesis, in: Zooming in on
  Axions in the Early Universe, CERN, 2020.

\bibitem{Harigaya2022Conf}
K.~Harigaya,
  \href{https://indico.in2p3.fr/event/24773/contributions/110607/attachments/71150/101092/2022Planck.pdf}{Cosmology
  of axion rotation}, Presentation at {Planck} conference 02/06/2022 (2022).
\newline\urlprefix\url{https://indico.in2p3.fr/event/24773/contributions/110607/attachments/71150/101092/2022Planck.pdf}

\bibitem{Harigaya2023UCLAConf}
K.~Harigaya,
  \href{https://indico.cern.ch/event/1188759/contributions/5244394/attachments/2621055/4533371/Harigaya%20(1).pdf}{Cosmology
  of axion rotation}, Presentation at {UCLA Dark Matter 2023} symposium 30
  March 2023 (2023).
\newline\urlprefix\url{https://indico.cern.ch/event/1188759/contributions/5244394/attachments/2621055/4533371/Harigaya%20(1).pdf}

\bibitem{Co2020Misalignment}
R.~T. Co, L.~J. Hall, K.~Harigaya, Axion kinetic misalignment mechanism, Phys.
  Rev. Lett. 124 (2020) 251802.
\newblock \href {https://doi.org/10.1103/PhysRevLett.124.251802}
  {\path{doi:10.1103/PhysRevLett.124.251802}}.

\bibitem{Ellis1987}
J.~R. Ellis, K.~A. Olive, {Constraints on Light Particles From Supernova
  Sn1987a}, Phys. Lett. B 193 (1987) 525.
\newblock \href {https://doi.org/10.1016/0370-2693(87)91710-2}
  {\path{doi:10.1016/0370-2693(87)91710-2}}.

\bibitem{Raffelt1988SN1987A}
G.~Raffelt, D.~Seckel, Bounds on exotic-particle interactions from sn1987a,
  Phys. Rev. Lett. 60 (1988) 1793--1796.
\newblock \href {https://doi.org/10.1103/PhysRevLett.60.1793}
  {\path{doi:10.1103/PhysRevLett.60.1793}}.

\bibitem{Turner1988}
M.~S. Turner, Axions from sn1987a, Phys. Rev. Lett. 60 (1988) 1797--1800.
\newblock \href {https://doi.org/10.1103/PhysRevLett.60.1797}
  {\path{doi:10.1103/PhysRevLett.60.1797}}.

\bibitem{Mayle1988}
R.~Mayle, J.~R. Wilson, J.~Ellis, K.~Olive, D.~N. Schramm, G.~Steigman,
  Constraints on axions from sn 1987a, Physics Letters B 203~(1) (1988)
  188--196.
\newblock \href {https://doi.org/https://doi.org/10.1016/0370-2693(88)91595-X}
  {\path{doi:https://doi.org/10.1016/0370-2693(88)91595-X}}.

\bibitem{Payez2015}
A.~Payez, C.~Evoli, T.~Fischer, M.~Giannotti, A.~Mirizzi, A.~Ringwald,
  Revisiting the {SN}1987a gamma-ray limit on ultralight axion-like particles,
  Journal of Cosmology and Astroparticle Physics 2015~(02) (2015) 006--006.
\newblock \href {https://doi.org/10.1088/1475-7516/2015/02/006}
  {\path{doi:10.1088/1475-7516/2015/02/006}}.

\bibitem{Bar2020}
N.~Bar, K.~Blum, G.~D'Amico, Is there a supernova bound on axions?, Phys. Rev.
  D 101 (2020) 123025.
\newblock \href {https://doi.org/10.1103/PhysRevD.101.123025}
  {\path{doi:10.1103/PhysRevD.101.123025}}.

\bibitem{DOnofrio2014}
M.~D'Onofrio, K.~Rummukainen, A.~Tranberg, Sphaleron rate in the minimal
  standard model, Phys. Rev. Lett. 113 (2014) 141602.
\newblock \href {https://doi.org/10.1103/PhysRevLett.113.141602}
  {\path{doi:10.1103/PhysRevLett.113.141602}}.

\bibitem{Chang1993}
S.~Chang, K.~Choi, Hadronic axion window and the big-bang nucleosynthesis,
  Physics Letters B 316~(1) (1993) 51--56.
\newblock \href {https://doi.org/10.1016/0370-2693(93)90656-3}
  {\path{doi:10.1016/0370-2693(93)90656-3}}.

\bibitem{Merkotan2017}
K.~K. {Merkotan}, T.~M. {Zelentsova}, N.~O. {Chudak}, D.~A. {Ptashynskyi},
  V.~V. {Urbanevich}, O.~S. {Potiienko}, V.~V. {Voitenko}, O.~D. {Berezovskyi},
  I.~V. {Sharph}, V.~D. {Rusov}, {An Alternative Method for Solving Two
  Problems of the Standard Model}, ArXiv e-prints (Oct. 2017).
\newblock \href {http://arxiv.org/abs/1711.01914} {\path{arXiv:1711.01914}},
  \href {https://doi.org/10.48550/arXiv.1711.01914}
  {\path{doi:10.48550/arXiv.1711.01914}}.

\bibitem{Merkotan2018}
K.~K. Merkotan, T.~M. Zelentsova, N.~O. Chudak, D.~A. Ptashynskiy, V.~V.
  Urbanevich, O.~S. Potiienko, V.~V. Voitenko, O.~D. Berezovskyi, I.~V. Sharph,
  V.~D. Rusov, Multi-particle fields and higgs mechanism, Journal of Physical
  Studies 22~(3), (in Ukrainian) (2018).
\newblock \href {https://doi.org/10.30970/jps.22.3001}
  {\path{doi:10.30970/jps.22.3001}}.

\bibitem{Ptashynskiy2019}
D.~A. Ptashynskiy, T.~M. Zelentsova, N.~O. Chudak, K.~K. Merkotan, O.~S.
  Potiienko, V.~V. Voitenko, O.~D. Berezovskiy, V.~V. Opyatyuk, O.~V. Zharova,
  T.~V. Yushkevich, I.~V. Sharph, V.~D. Rusov, Multiparticle fields on the
  subset of simultaneity, Ukrainian Journal of Physics 64~(8) (2019) 732.
\newblock \href {https://doi.org/10.15407/ujpe64.8.732}
  {\path{doi:10.15407/ujpe64.8.732}}.

\bibitem{Merkotan2021PhD}
K.~K. Merkotan, Application of multiparticle fields in the theory of
  electroweak interaction, Ph.D. thesis, Odessa National Polytechnic
  University, Odessa, Ukraine (2021).

\bibitem{Li2020}
X.~Li, A.~Shafieloo, Evidence for emergent dark energy, The Astrophysical
  Journal 902~(1) (2020) 58.
\newblock \href {https://doi.org/10.3847/1538-4357/abb3d0}
  {\path{doi:10.3847/1538-4357/abb3d0}}.

\bibitem{Li2019}
X.~Li, A.~Shafieloo, A simple phenomenological emergent dark energy model can
  resolve the hubble tension, The Astrophysical Journal 883~(1) (2019) L3.
\newblock \href {https://doi.org/10.3847/2041-8213/ab3e09}
  {\path{doi:10.3847/2041-8213/ab3e09}}.

\bibitem{Jeans1902}
J.~H. Jeans, The stability of a spherical nebula, Philosophical Transactions of
  the Royal Society of London. Series A, Containing Papers of a Mathematical or
  Physical Character 199 (1902) 1--53.

\bibitem{Jeans1928}
J.~Jeans, Astronomy and Cosmogony, 2nd Edition, Cambridge Library Collection -
  Astronomy, Cambridge University Press, 1928.

\bibitem{Yang2020}
W.~Yang, H.~Chen, S.~Liu, {The effect of dark matter on the Jeans instability
  with the q-nonextensive velocity distribution}, AIP Advances 10~(7) (2020)
  075003.
\newblock \href {https://doi.org/10.1063/5.0011567}
  {\path{doi:10.1063/5.0011567}}.

\bibitem{Kremer2016}
G.~M. Kremer, R.~Andr\'{e}, Analysis of instability of systems composed by dark
  and baryonic matter, International Journal of Modern Physics D 25~(01) (2016)
  1650012.
\newblock \href {https://doi.org/10.1142/S0218271816500127}
  {\path{doi:10.1142/S0218271816500127}}.

\bibitem{Kremer2018}
G.~M. Kremer, M.~G. Richarte, F.~Teston, Jeans instability in a universe with
  dissipation, Phys. Rev. D 97 (2018) 023515.
\newblock \href {https://doi.org/10.1103/PhysRevD.97.023515}
  {\path{doi:10.1103/PhysRevD.97.023515}}.

\bibitem{Kremer2019}
G.~M. Kremer, M.~G. Richarte, E.~M. Schiefer, Using kinetic theory to examine a
  self-gravitating system composed of baryons and cold dark matter, The
  European Physical Journal C 79 (2019) 1--11.

\bibitem{Snell1981}
R.~L. {Snell}, {A study of nine interstellar dark clouds.}, The Astrophysical
  Journal Supplement 45 (1981) 121--175.
\newblock \href {https://doi.org/10.1086/190711} {\path{doi:10.1086/190711}}.

\bibitem{Dwivedi1999}
C.~B. {Dwivedi}, A.~K. {Sen}, S.~{Bujarbarua}, {Pulsational mode of
  gravitational collapse and its impact on the star formation}, Astronomy and
  Astrophysics 345 (1999) 1049--1053.

\bibitem{Prialnik2009}
D.~Prialnik, An Introduction to the Theory of Stellar Structure and Evolution,
  Cambridge University Press, 2009.

\bibitem{Anselm1982}
A.~Anselm, N.~Uraltsev, Long range “arion” field in the radio frequency
  band, Physics Letters B 116~(2) (1982) 161--164.
\newblock \href {https://doi.org/10.1016/0370-2693(82)90999-6}
  {\path{doi:10.1016/0370-2693(82)90999-6}}.

\bibitem{Vincent2020}
A.~C. Vincent, Dark matter in stars (2020).
\newblock \href {http://arxiv.org/abs/2009.00663} {\path{arXiv:2009.00663}},
  \href {https://doi.org/10.48550/arXiv.2009.00663}
  {\path{doi:10.48550/arXiv.2009.00663}}.

\bibitem{Dearborn1990}
D.~Dearborn, G.~Raffelt, P.~Salati, J.~Silk, A.~Bouquet, Dark matter and
  thermal pulses in horizontal-branch stars, Astrophysical Journal 354 (5
  1990).
\newblock \href {https://doi.org/10.1086/168716} {\path{doi:10.1086/168716}}.

\bibitem{Spergel1985}
D.~N. {Spergel}, W.~H. {Press}, {Effect of hypothetical, weakly interacting,
  massive particles on energy transport in the solar interior}, Astrophysical
  Journal 294 (1985) 663--673.
\newblock \href {https://doi.org/10.1086/163336} {\path{doi:10.1086/163336}}.

\bibitem{GouldRaffelt1990a}
A.~{Gould}, G.~{Raffelt}, {Thermal conduction by massive particles},
  Astrophysical Journal 352 (1990) 654--668.
\newblock \href {https://doi.org/10.1086/168568} {\path{doi:10.1086/168568}}.

\bibitem{GouldRaffelt1990b}
A.~{Gould}, G.~{Raffelt}, {Cosmion energy transfer in stars - The Knudsen
  limit}, Astrophysical Journal 352 (1990) 669--680.
\newblock \href {https://doi.org/10.1086/168569} {\path{doi:10.1086/168569}}.

\bibitem{Raffelt1986}
G.~G. Raffelt, Astrophysical axion bounds diminished by screening effects,
  Phys. Rev. D 33 (1986) 897--909.
\newblock \href {https://doi.org/10.1103/PhysRevD.33.897}
  {\path{doi:10.1103/PhysRevD.33.897}}.

\bibitem{Connelly2012}
J.~N. Connelly, M.~Bizzarro, A.~N. Krot, Åke Nordlund, D.~Wielandt, M.~A.
  Ivanova, The absolute chronology and thermal processing of solids in the
  solar protoplanetary disk, Science 338~(6107) (2012) 651--655.
\newblock \href {https://doi.org/10.1126/science.1226919}
  {\path{doi:10.1126/science.1226919}}.

\bibitem{Subr2019}
L.~\v{S}ubr, G.~Fragione, J.~Dabringhausen, {Intermediate-mass black holes in
  binary-rich star clusters}, Monthly Notices of the Royal Astronomical Society
  484~(3) (2019) 2974--2986.
\newblock \href {https://doi.org/10.1093/mnras/stz162}
  {\path{doi:10.1093/mnras/stz162}}.

\bibitem{Vincent2016}
A.~C. {Vincent}, P.~{Scott}, A.~{Serenelli}, {Updated constraints on velocity
  and momentum-dependent asymmetric dark matter}, Journal of Cosmology and
  Astroparticle Physics 11 (2016) 007.
\newblock \href {http://arxiv.org/abs/1605.06502} {\path{arXiv:1605.06502}},
  \href {https://doi.org/10.1088/1475-7516/2016/11/007}
  {\path{doi:10.1088/1475-7516/2016/11/007}}.

\bibitem{Grevesse1998}
N.~{Grevesse}, A.~J. {Sauval}, {Standard Solar Composition}, Space Science
  Reviews 85 (1998) 161--174.
\newblock \href {https://doi.org/10.1023/A:1005161325181}
  {\path{doi:10.1023/A:1005161325181}}.

\bibitem{Serenelli2013}
A.~Serenelli, C.~Pe\~na Garay, W.~C. Haxton, Using the standard solar model to
  constrain solar composition and nuclear reaction $s$ factors, Phys. Rev. D 87
  (2013) 043001.
\newblock \href {https://doi.org/10.1103/PhysRevD.87.043001}
  {\path{doi:10.1103/PhysRevD.87.043001}}.

\bibitem{Villante2014}
F.~L. {Villante}, A.~M. {Serenelli}, F.~{Delahaye}, M.~H. {Pinsonneault}, {The
  Chemical Composition of the Sun from Helioseismic and Solar Neutrino Data},
  Astrophysical Journal 787 (2014) 13.
\newblock \href {http://arxiv.org/abs/1312.3885} {\path{arXiv:1312.3885}},
  \href {https://doi.org/10.1088/0004-637X/787/1/13}
  {\path{doi:10.1088/0004-637X/787/1/13}}.

\bibitem{Vinyoles2015}
N.~Vinyoles, A.~Serenelli, F.~L. Villante, S.~Basu, J.~Redondo, J.~Isern, New
  axion and hidden photon constraints from a solar data global fit, Journal of
  Cosmology and Astroparticle Physics 10 (2015) 015.
\newblock \href {http://arxiv.org/abs/1501.01639} {\path{arXiv:1501.01639}},
  \href {https://doi.org/10.1088/1475-7516/2015/10/015}
  {\path{doi:10.1088/1475-7516/2015/10/015}}.

\bibitem{DeglInnoccenti1997}
S.~{Degl'Innoccenti}, W.~A. {Dziembowski}, G.~{Fiorentini}, B.~{Ricci},
  {Helioseismology and standard solar models}, Astroparticle Physics 7 (1997)
  77--95.
\newblock \href {http://arxiv.org/abs/astro-ph/9612053}
  {\path{arXiv:astro-ph/9612053}}, \href
  {https://doi.org/10.1016/S0927-6505(97)00004-2}
  {\path{doi:10.1016/S0927-6505(97)00004-2}}.

\bibitem{Fiorentini2001}
G.~Fiorentini, B.~Ricci, F.~Villante, Helioseismology and screening of nuclear
  reactions in the sun, Physics Letters B 503~(1) (2001) 121 -- 125.
\newblock \href {https://doi.org/10.1016/S0370-2693(01)00221-0}
  {\path{doi:10.1016/S0370-2693(01)00221-0}}.

\bibitem{Goulding2018}
A.~D. {Goulding}, J.~E. {Greene}, R.~{Bezanson}, J.~{Greco}, S.~{Johnson},
  A.~{Leauthaud}, Y.~{Matsuoka}, E.~{Medezinski}, A.~M. {Price-Whelan}, {Galaxy
  interactions trigger rapid black hole growth: An unprecedented view from the
  Hyper Suprime-Cam survey}, Publications of the Astronomical Society of Japan
  70 (2018) S37.
\newblock \href {http://arxiv.org/abs/1706.07436} {\path{arXiv:1706.07436}},
  \href {https://doi.org/10.1093/pasj/psx135} {\path{doi:10.1093/pasj/psx135}}.

\bibitem{Rusov2015}
V.~Rusov, M.~Eingorn, I.~Sharph, V.~Smolyar, M.~Beglaryan, Thermomagnetic
  {Ettingshausen-Nernst} effect in tachocline and axion mechanism of solar
  luminosity variations,
  \href{https://arxiv.org/abs/1508.03836}{arXiv:1508.03836 [astro-ph.SR]}
  (2015).

\bibitem{Volkotrub2015}
Y.~V. {Volkotrub}, M.~A. {Deliyergiyev}, K.~K. {Merkotan}, N.~A. {Chudak},
  O.~S. {Potiyenko}, D.~A. {Ptashynskyy}, G.~O. {Sokhrannyi}, A.~V. {Tykhonov},
  Y.~V. {Shabatura}, I.~V. {Sharph}, V.~D. {Rusov}, {Multi-particle field
  operators in quantum field theory}, ArXiv e-prints (Sep. 2015).
\newblock \href {http://arxiv.org/abs/1510.01937} {\path{arXiv:1510.01937}},
  \href {https://doi.org/10.48550/arXiv.1510.01937}
  {\path{doi:10.48550/arXiv.1510.01937}}.

\bibitem{Chudak2016}
N.~A. {Chudak}, M.~A. {Deliyergiyev}, K.~K. {Merkotan}, O.~S. {Potiienko},
  D.~A. {Ptashynskyi}, Y.~V. {Shabatura}, G.~O. {Sokhrannyi}, A.~V. {Tykhonov},
  Y.~V. {Volkotrub}, I.~V. {Sharph}, V.~D. {Rusov}, {Multiparticle quantum
  fields}, Physics Journal (2016) 181.

\bibitem{CMS2020}
{The CMS Collaboration}, A measurement of the higgs boson mass in the diphoton
  decay channel, Physics Letters B 805 (2020) 135425.
\newblock \href {https://doi.org/10.1016/j.physletb.2020.135425}
  {\path{doi:10.1016/j.physletb.2020.135425}}.

\bibitem{CMS2022a}
{The {CMS} Collaboration}, A portrait of the higgs boson by the {CMS}
  experiment ten years after the discovery, Nature 607~(7917) (2022) 60–68.
\newblock \href {https://doi.org/10.1038/s41586-022-04892-x}
  {\path{doi:10.1038/s41586-022-04892-x}}.

\bibitem{CMS2022b}
{The {CMS} Collaboration}, Measurement of the higgs boson width and evidence of
  its off-shell contributions to {ZZ} production, Nature Physics 18~(11) (2022)
  1329--1334.
\newblock \href {https://doi.org/10.1038/s41567-022-01682-0}
  {\path{doi:10.1038/s41567-022-01682-0}}.

\bibitem{Planck2015}
{Planck Collaboration}, {Ade, P. A. R.}, {Aghanim, N.}, {Arnaud, M.}, {Ashdown,
  M.}, {Aumont, J.}, {Baccigalupi, C.}, {Banday, A. J.}, {Barreiro, R. B.},
  {Bartlett, J. G.}, {Bartolo, N.}, {Battaner, E.}, {Battye, R.}, {Benabed,
  K.}, {Beno\^{\i}t, A.}, {Benoit-L\'evy, A.}, {Bernard, J.-P.}, {Bersanelli,
  M.}, {Bielewicz, P.}, {Bock, J. J.}, {Bonaldi, A.}, {Bonavera, L.}, {Bond, J.
  R.}, {Borrill, J.}, {Bouchet, F. R.}, {Boulanger, F.}, {Bucher, M.},
  {Burigana, C.}, {Butler, R. C.}, {Calabrese, E.}, {Cardoso, J.-F.},
  {Catalano, A.}, {Challinor, A.}, {Chamballu, A.}, {Chary, R.-R.}, {Chiang, H.
  C.}, {Chluba, J.}, {Christensen, P. R.}, {Church, S.}, {Clements, D. L.},
  {Colombi, S.}, {Colombo, L. P. L.}, {Combet, C.}, {Coulais, A.}, {Crill, B.
  P.}, {Curto, A.}, {Cuttaia, F.}, {Danese, L.}, {Davies, R. D.}, {Davis, R.
  J.}, {de Bernardis, P.}, {de Rosa, A.}, {de Zotti, G.}, {Delabrouille, J.},
  {D\'esert, F.-X.}, {Di Valentino, E.}, {Dickinson, C.}, {Diego, J. M.},
  {Dolag, K.}, {Dole, H.}, {Donzelli, S.}, {Dor\'e, O.}, {Douspis, M.},
  {Ducout, A.}, {Dunkley, J.}, {Dupac, X.}, {Efstathiou, G.}, {Elsner, F.},
  {En\ss{}lin, T. A.}, {Eriksen, H. K.}, {Farhang, M.}, {Fergusson, J.},
  {Finelli, F.}, {Forni, O.}, {Frailis, M.}, {Fraisse, A. A.}, {Franceschi,
  E.}, {Frejsel, A.}, {Galeotta, S.}, {Galli, S.}, {Ganga, K.}, {Gauthier, C.},
  {Gerbino, M.}, {Ghosh, T.}, {Giard, M.}, {Giraud-H\'eraud, Y.}, {Giusarma,
  E.}, {Gjerl\o{}w, E.}, {Gonz\'alez-Nuevo, J.}, {G\'orski, K. M.}, {Gratton,
  S.}, {Gregorio, A.}, {Gruppuso, A.}, {Gudmundsson, J. E.}, {Hamann, J.},
  {Hansen, F. K.}, {Hanson, D.}, {Harrison, D. L.}, {Helou, G.},
  {Henrot-Versill\'e, S.}, {Hern\'andez-Monteagudo, C.}, {Herranz, D.},
  {Hildebrandt, S. R.}, {Hivon, E.}, {Hobson, M.}, {Holmes, W. A.}, {Hornstrup,
  A.}, {Hovest, W.}, {Huang, Z.}, {Huffenberger, K. M.}, {Hurier, G.}, {Jaffe,
  A. H.}, {Jaffe, T. R.}, {Jones, W. C.}, {Juvela, M.}, {Keih\"anen, E.},
  {Keskitalo, R.}, {Kisner, T. S.}, {Kneissl, R.}, {Knoche, J.}, {Knox, L.},
  {Kunz, M.}, {Kurki-Suonio, H.}, {Lagache, G.}, {L\"ahteenm\"aki, A.},
  {Lamarre, J.-M.}, {Lasenby, A.}, {Lattanzi, M.}, {Lawrence, C. R.}, {Leahy,
  J. P.}, {Leonardi, R.}, {Lesgourgues, J.}, {Levrier, F.}, {Lewis, A.},
  {Liguori, M.}, {Lilje, P. B.}, {Linden-V\o{}rnle, M.}, {L\'opez-Caniego, M.},
  {Lubin, P. M.}, {Mac\'{\i}as-P\'erez, J. F.}, {Maggio, G.}, {Maino, D.},
  {Mandolesi, N.}, {Mangilli, A.}, {Marchini, A.}, {Maris, M.}, {Martin, P.
  G.}, {Martinelli, M.}, {Mart\'{\i}nez-Gonz\'alez, E.}, {Masi, S.},
  {Matarrese, S.}, {McGehee, P.}, {Meinhold, P. R.}, {Melchiorri, A.}, {Melin,
  J.-B.}, {Mendes, L.}, {Mennella, A.}, {Migliaccio, M.}, {Millea, M.}, {Mitra,
  S.}, {Miville-Desch\^enes, M.-A.}, {Moneti, A.}, {Montier, L.}, {Morgante,
  G.}, {Mortlock, D.}, {Moss, A.}, {Munshi, D.}, {Murphy, J. A.}, {Naselsky,
  P.}, {Nati, F.}, {Natoli, P.}, {Netterfield, C. B.}, {N\o{}rgaard-Nielsen, H.
  U.}, {Noviello, F.}, {Novikov, D.}, {Novikov, I.}, {Oxborrow, C. A.}, {Paci,
  F.}, {Pagano, L.}, {Pajot, F.}, {Paladini, R.}, {Paoletti, D.}, {Partridge,
  B.}, {Pasian, F.}, {Patanchon, G.}, {Pearson, T. J.}, {Perdereau, O.},
  {Perotto, L.}, {Perrotta, F.}, {Pettorino, V.}, {Piacentini, F.}, {Piat, M.},
  {Pierpaoli, E.}, {Pietrobon, D.}, {Plaszczynski, S.}, {Pointecouteau, E.},
  {Polenta, G.}, {Popa, L.}, {Pratt, G. W.}, {Pr\'ezeau, G.}, {Prunet, S.},
  {Puget, J.-L.}, {Rachen, J. P.}, {Reach, W. T.}, {Rebolo, R.}, {Reinecke,
  M.}, {Remazeilles, M.}, {Renault, C.}, {Renzi, A.}, {Ristorcelli, I.},
  {Rocha, G.}, {Rosset, C.}, {Rossetti, M.}, {Roudier, G.}, {Rouill\'e
  d\'{}Orfeuil, B.}, {Rowan-Robinson, M.}, {Rubi\~no-Mart\'{\i}n, J. A.},
  {Rusholme, B.}, {Said, N.}, {Salvatelli, V.}, {Salvati, L.}, {Sandri, M.},
  {Santos, D.}, {Savelainen, M.}, {Savini, G.}, {Scott, D.}, {Seiffert, M. D.},
  {Serra, P.}, {Shellard, E. P. S.}, {Spencer, L. D.}, {Spinelli, M.},
  {Stolyarov, V.}, {Stompor, R.}, {Sudiwala, R.}, {Sunyaev, R.}, {Sutton, D.},
  {Suur-Uski, A.-S.}, {Sygnet, J.-F.}, {Tauber, J. A.}, {Terenzi, L.},
  {Toffolatti, L.}, {Tomasi, M.}, {Tristram, M.}, {Trombetti, T.}, {Tucci, M.},
  {Tuovinen, J.}, {T\"urler, M.}, {Umana, G.}, {Valenziano, L.}, {Valiviita,
  J.}, {Van Tent, F.}, {Vielva, P.}, {Villa, F.}, {Wade, L. A.}, {Wandelt, B.
  D.}, {Wehus, I. K.}, {White, M.}, {White, S. D. M.}, {Wilkinson, A.}, {Yvon,
  D.}, {Zacchei, A.}, {Zonca, A.}, Planck 2015 results - xiii. cosmological
  parameters, A\&A 594 (2016) A13.
\newblock \href {https://doi.org/10.1051/0004-6361/201525830}
  {\path{doi:10.1051/0004-6361/201525830}}.

\bibitem{Velten2014}
H.~E.~S. Velten, R.~F. vom Marttens, W.~Zimdahl, Aspects of the cosmological
  “coincidence problem”, The European Physical Journal C 74~(11) (2014)
  3160.
\newblock \href {https://doi.org/10.1140/epjc/s10052-014-3160-4}
  {\path{doi:10.1140/epjc/s10052-014-3160-4}}.

\bibitem{Nussinov1985}
S.~Nussinov, Technocosmology -- could a technibaryon excess provide a
  “natural” missing mass candidate?, Physics Letters B 165~(1–3) (1985)
  55–58.
\newblock \href {https://doi.org/10.1016/0370-2693(85)90689-6}
  {\path{doi:10.1016/0370-2693(85)90689-6}}.

\bibitem{Chivukula1990}
R.~Chivukula, T.~P. Walker, Technicolor cosmology, Nuclear Physics B 329~(2)
  (1990) 445–463.
\newblock \href {https://doi.org/10.1016/0550-3213(90)90151-3}
  {\path{doi:10.1016/0550-3213(90)90151-3}}.

\bibitem{Barr1990}
S.~Barr, R.~Sekhar~Chivukula, E.~Farhi, Electroweak fermion number violation
  and the production of stable particles in the early universe, Physics Letters
  B 241~(3) (1990) 387–391.
\newblock \href {https://doi.org/10.1016/0370-2693(90)91661-T}
  {\path{doi:10.1016/0370-2693(90)91661-T}}.

\bibitem{Kaplan1992}
D.~B. Kaplan, Single explanation for both baryon and dark matter densities,
  Physical Review Letters 68~(6) (1992) 741–743.
\newblock \href {https://doi.org/10.1103/PhysRevLett.68.741}
  {\path{doi:10.1103/PhysRevLett.68.741}}.

\bibitem{Hooper2005}
D.~Hooper, J.~March-Russell, S.~M. West, Asymmetric sneutrino dark matter and
  the $\omega_b$/$\omega_{DM}$ puzzle, Physics Letters B 605~(3–4) (2005)
  228–236.
\newblock \href {https://doi.org/10.1016/j.physletb.2004.11.047}
  {\path{doi:10.1016/j.physletb.2004.11.047}}.

\bibitem{Kaplan2009}
D.~E. Kaplan, M.~A. Luty, K.~M. Zurek, Asymmetric dark matter, Physical Review
  D 79~(11) (2009) 115016.
\newblock \href {https://doi.org/10.1103/PhysRevD.79.115016}
  {\path{doi:10.1103/PhysRevD.79.115016}}.

\bibitem{Davoudiasl2012}
H.~Davoudiasl, R.~N. Mohapatra, On relating the genesis of cosmic baryons and
  dark matter, New Journal of Physics 14~(9) (2012) 095011.
\newblock \href {https://doi.org/10.1088/1367-2630/14/9/095011}
  {\path{doi:10.1088/1367-2630/14/9/095011}}.

\bibitem{Petraki2013}
K.~{Petraki}, R.~R. {Volkas}, {Review of Asymmetric Dark Matter}, International
  Journal of Modern Physics A 28 (2013) 1330028.
\newblock \href {http://arxiv.org/abs/1305.4939} {\path{arXiv:1305.4939}},
  \href {https://doi.org/10.1142/S0217751X13300287}
  {\path{doi:10.1142/S0217751X13300287}}.

\bibitem{Zurek2014}
K.~M. {Zurek}, {Asymmetric Dark Matter: Theories, signatures, and constraints},
  Physics Reports 537 (2014) 91--121.
\newblock \href {http://arxiv.org/abs/1308.0338} {\path{arXiv:1308.0338}},
  \href {https://doi.org/10.1016/j.physrep.2013.12.001}
  {\path{doi:10.1016/j.physrep.2013.12.001}}.

\bibitem{Beekman2019}
A.~J. Beekman, L.~Rademaker, J.~van Wezel, {An introduction to spontaneous
  symmetry breaking}, SciPost Phys. Lect. Notes (2019) 11\href
  {https://doi.org/10.21468/SciPostPhysLectNotes.11}
  {\path{doi:10.21468/SciPostPhysLectNotes.11}}.

\bibitem{CERNHiggs2022}
CERN, The {Higgs} boson, ten years after its discovery,
  https://home.cern/news/press-release/physics/higgs-boson-ten-years-after-its-discovery
  (4 July 2022).

\bibitem{Grojean2007}
C.~{Grojean}, {REVIEWS OF TOPICAL PROBLEMS: New approaches to electroweak
  symmetry breaking}, Physics Uspekhi 50 (2007) 1--35.
\newblock \href {https://doi.org/10.1070/PU2007v050n01ABEH006157}
  {\path{doi:10.1070/PU2007v050n01ABEH006157}}.

\bibitem{Salam2022}
G.~P. Salam, L.-T. Wang, G.~Zanderighi, The higgs boson turns ten, Nature
  607~(7917) (2022) 41–47.
\newblock \href {https://doi.org/10.1038/s41586-022-04899-4}
  {\path{doi:10.1038/s41586-022-04899-4}}.

\bibitem{Englert1964}
F.~Englert, R.~Brout, Broken symmetry and the mass of gauge vector mesons,
  Phys. Rev. Lett. 13 (1964) 321--323.
\newblock \href {https://doi.org/10.1103/PhysRevLett.13.321}
  {\path{doi:10.1103/PhysRevLett.13.321}}.

\bibitem{Higgs1964b}
P.~W. {Higgs}, {Broken Symmetries and the Masses of Gauge Bosons}, Physical
  Review Letters 13 (1964) 508--509.
\newblock \href {https://doi.org/10.1103/PhysRevLett.13.508}
  {\path{doi:10.1103/PhysRevLett.13.508}}.

\bibitem{Guralnik1964}
G.~S. Guralnik, C.~R. Hagen, T.~W.~B. Kibble, Global conservation laws and
  massless particles, Phys. Rev. Lett. 13 (1964) 585--587.
\newblock \href {https://doi.org/10.1103/PhysRevLett.13.585}
  {\path{doi:10.1103/PhysRevLett.13.585}}.

\bibitem{Weinberg1967}
S.~{Weinberg}, {A Model of Leptons}, Physical Review Letters 19 (1967)
  1264--1266.
\newblock \href {https://doi.org/10.1103/PhysRevLett.19.1264}
  {\path{doi:10.1103/PhysRevLett.19.1264}}.

\bibitem{Salam1964}
A.~{Salam}, J.~C. {Ward}, {Electromagnetic and weak interactions}, Physics
  Letters 13 (1964) 168--171.
\newblock \href {https://doi.org/10.1016/0031-9163(64)90711-5}
  {\path{doi:10.1016/0031-9163(64)90711-5}}.

\bibitem{Glashow1961}
S.~L. {Glashow}, {Partial-symmetries of weak interactions}, Nuclear Physics A
  22 (1961) 579--588.
\newblock \href {https://doi.org/10.1016/0029-5582(61)90469-2}
  {\path{doi:10.1016/0029-5582(61)90469-2}}.

\bibitem{TanabashiPDG2018}
M.~Tanabashi, K.~Hagiwara, K.~Hikasa, K.~Nakamura, Y.~Sumino, F.~Takahashi,
  J.~Tanaka, K.~Agashe, G.~Aielli, C.~Amsler, M.~Antonelli, D.~M. Asner,
  H.~Baer, S.~Banerjee, R.~M. Barnett, T.~Basaglia, C.~W. Bauer, J.~J. Beatty,
  V.~I. Belousov, J.~Beringer, S.~Bethke, A.~Bettini, H.~Bichsel, O.~Biebel,
  K.~M. Black, E.~Blucher, O.~Buchmuller, V.~Burkert, M.~A. Bychkov, R.~N.
  Cahn, M.~Carena, A.~Ceccucci, A.~Cerri, D.~Chakraborty, M.-C. Chen, R.~S.
  Chivukula, G.~Cowan, O.~Dahl, G.~D'Ambrosio, T.~Damour, D.~de~Florian,
  A.~de~Gouv\^ea, T.~DeGrand, P.~de~Jong, G.~Dissertori, B.~A. Dobrescu,
  M.~D'Onofrio, M.~Doser, M.~Drees, H.~K. Dreiner, D.~A. Dwyer, P.~Eerola,
  S.~Eidelman, J.~Ellis, J.~Erler, V.~V. Ezhela, W.~Fetscher, B.~D. Fields,
  R.~Firestone, B.~Foster, A.~Freitas, H.~Gallagher, L.~Garren, H.-J. Gerber,
  G.~Gerbier, T.~Gershon, Y.~Gershtein, T.~Gherghetta, A.~A. Godizov,
  M.~Goodman, C.~Grab, A.~V. Gritsan, C.~Grojean, D.~E. Groom, M.~Gr\"unewald,
  A.~Gurtu, T.~Gutsche, H.~E. Haber, C.~Hanhart, S.~Hashimoto, Y.~Hayato, K.~G.
  Hayes, A.~Hebecker, S.~Heinemeyer, B.~Heltsley, J.~J. Hern\'andez-Rey,
  J.~Hisano, A.~H\"ocker, J.~Holder, A.~Holtkamp, T.~Hyodo, K.~D. Irwin, K.~F.
  Johnson, M.~Kado, M.~Karliner, U.~F. Katz, S.~R. Klein, E.~Klempt, R.~V.
  Kowalewski, F.~Krauss, M.~Kreps, B.~Krusche, Y.~V. Kuyanov, Y.~Kwon,
  O.~Lahav, J.~Laiho, J.~Lesgourgues, A.~Liddle, Z.~Ligeti, C.-J. Lin,
  C.~Lippmann, T.~M. Liss, L.~Littenberg, K.~S. Lugovsky, S.~B. Lugovsky,
  A.~Lusiani, Y.~Makida, F.~Maltoni, T.~Mannel, A.~V. Manohar, W.~J. Marciano,
  A.~D. Martin, A.~Masoni, J.~Matthews, U.-G. Mei\ss{}ner, D.~Milstead, R.~E.
  Mitchell, K.~M\"onig, P.~Molaro, F.~Moortgat, M.~Moskovic, H.~Murayama,
  M.~Narain, P.~Nason, S.~Navas, M.~Neubert, P.~Nevski, Y.~Nir, K.~A. Olive,
  S.~Pagan~Griso, J.~Parsons, C.~Patrignani, J.~A. Peacock, M.~Pennington,
  S.~T. Petcov, V.~A. Petrov, E.~Pianori, A.~Piepke, A.~Pomarol, A.~Quadt,
  J.~Rademacker, G.~Raffelt, B.~N. Ratcliff, P.~Richardson, A.~Ringwald,
  S.~Roesler, S.~Rolli, A.~Romaniouk, L.~J. Rosenberg, J.~L. Rosner, G.~Rybka,
  R.~A. Ryutin, C.~T. Sachrajda, Y.~Sakai, G.~P. Salam, S.~Sarkar, F.~Sauli,
  O.~Schneider, K.~Scholberg, A.~J. Schwartz, D.~Scott, V.~Sharma, S.~R.
  Sharpe, T.~Shutt, M.~Silari, T.~Sj\"ostrand, P.~Skands, T.~Skwarnicki, J.~G.
  Smith, G.~F. Smoot, S.~Spanier, H.~Spieler, C.~Spiering, A.~Stahl, S.~L.
  Stone, T.~Sumiyoshi, M.~J. Syphers, K.~Terashi, J.~Terning, U.~Thoma, R.~S.
  Thorne, L.~Tiator, M.~Titov, N.~P. Tkachenko, N.~A. T\"ornqvist, D.~R. Tovey,
  G.~Valencia, R.~Van~de Water, N.~Varelas, G.~Venanzoni, L.~Verde, M.~G.
  Vincter, P.~Vogel, A.~Vogt, S.~P. Wakely, W.~Walkowiak, C.~W. Walter,
  D.~Wands, D.~R. Ward, M.~O. Wascko, G.~Weiglein, D.~H. Weinberg, E.~J.
  Weinberg, M.~White, L.~R. Wiencke, S.~Willocq, C.~G. Wohl, J.~Womersley,
  C.~L. Woody, R.~L. Workman, W.-M. Yao, G.~P. Zeller, O.~V. Zenin, R.-Y. Zhu,
  S.-L. Zhu, F.~Zimmermann, P.~A. Zyla, J.~Anderson, L.~Fuller, V.~S. Lugovsky,
  P.~Schaffner, Review of particle physics, Phys. Rev. D 98 (2018) 030001.
\newblock \href {https://doi.org/10.1103/PhysRevD.98.030001}
  {\path{doi:10.1103/PhysRevD.98.030001}}.

\bibitem{ATLAS2022}
{The {ATLAS} Collaboration}, A detailed map of higgs boson interactions by the
  atlas experiment ten years after the discovery, Nature 607~(7917) (2022)
  52–59.
\newblock \href {https://doi.org/10.1038/s41586-022-04893-w}
  {\path{doi:10.1038/s41586-022-04893-w}}.

\bibitem{Yukawa1949}
H.~Yukawa, On the radius of the elementary particle, Phys. Rev. 76 (1949)
  300--301.
\newblock \href {https://doi.org/10.1103/PhysRev.76.300.2}
  {\path{doi:10.1103/PhysRev.76.300.2}}.

\bibitem{Yukawa1950a}
H.~Yukawa, Quantum theory of non-local fields. part {I}. free fields, Phys.
  Rev. 77 (1950) 219--226.
\newblock \href {https://doi.org/10.1103/PhysRev.77.219}
  {\path{doi:10.1103/PhysRev.77.219}}.

\bibitem{Yukawa1950b}
H.~Yukawa, Quantum theory of non-local fields. part {II}. irreducible fields
  and their interaction, Phys. Rev. 80 (1950) 1047--1052.
\newblock \href {https://doi.org/10.1103/PhysRev.80.1047}
  {\path{doi:10.1103/PhysRev.80.1047}}.

\bibitem{Dirac1949}
P.~A.~M. Dirac, Forms of relativistic dynamics, Rev. Mod. Phys. 21 (1949)
  392--399.
\newblock \href {https://doi.org/10.1103/RevModPhys.21.392}
  {\path{doi:10.1103/RevModPhys.21.392}}.

\bibitem{Heinzl2001}
T.~Heinzl, Light-Cone Quantization: Foundations and Applications, Springer
  Berlin Heidelberg, Berlin, Heidelberg, 2001, pp. 55--142.
\newblock \href {https://doi.org/10.1007/3-540-45114-5_2}
  {\path{doi:10.1007/3-540-45114-5_2}}.

\bibitem{Logunov1963a}
A.~A. Logunov, A.~N. Tavkhelidze, Quasi-optical approach in quantum field
  theory, Il Nuovo Cimento 29~(2) (1963) 380–399.
\newblock \href {https://doi.org/10.1007/BF02750359}
  {\path{doi:10.1007/BF02750359}}.

\bibitem{Logunov1963b}
A.~A. Logunov, A.~N. Tavkhelidze, I.~T. Todorov, O.~A. Khrustalev,
  Quasi-potential character of the scattering amplitude, Il Nuovo Cimento
  30~(1) (1963) 134–142.
\newblock \href {https://doi.org/10.1007/BF02750754}
  {\path{doi:10.1007/BF02750754}}.

\bibitem{Faustov1973}
R.~Faustov, Relativistic wavefunction and form factors of the bound system,
  Annals of Physics 78~(1) (1973) 176--189.
\newblock \href {https://doi.org/10.1016/0003-4916(73)90007-9}
  {\path{doi:10.1016/0003-4916(73)90007-9}}.

\bibitem{Tomonaga1946}
S.~Tomonaga, {On a Relativistically Invariant Formulation of the Quantum Theory
  of Wave Fields}, Progress of Theoretical Physics 1~(2) (1946) 27--42.
\newblock \href {https://doi.org/10.1143/PTP.1.27}
  {\path{doi:10.1143/PTP.1.27}}.

\bibitem{Dirac1932}
P.~Dirac, W.~Fock, B.~Podolsky, On quantum electrodynamics, Phys. Zs. Sowjet.
  (1932) 468.

\bibitem{Petrat2014a}
S.~Petrat, R.~Tumulka, Multi-time wave functions for quantum field theory,
  Annals of Physics 345 (2014) 17--54.
\newblock \href {https://doi.org/10.1016/j.aop.2014.03.004}
  {\path{doi:10.1016/j.aop.2014.03.004}}.

\bibitem{Chudak2016UJP}
N.~O. Chudak, K.~K. Merkotan, D.~A. Ptashynskyy, O.~S. Potiyenko, M.~A.
  Deliyergiyev, A.~V. Tikhonov, G.~O. Sokhrannyi, O.~V. Zharova, O.~D.
  Berezovs’kyi, V.~V. Voitenko, Y.~V. Volkotrub, I.~V. Sharph, V.~D. Rusov,
  Internal states of hadrons in relativistic reference frames, Ukrainian
  Journal of Physics 61~(12) (2016) 1033.
\newblock \href {https://doi.org/10.15407/ujpe61.12.1033}
  {\path{doi:10.15407/ujpe61.12.1033}}.

\bibitem{Marx1972}
E.~Marx, Generalized relativistic fock space, International Journal of
  Theoretical Physics 6~(5) (1972) 359–363.
\newblock \href {https://doi.org/10.1007/BF01258729}
  {\path{doi:10.1007/BF01258729}}.

\bibitem{Sazdjian1986}
H.~Sazdjian, Relativistic wave equations for the dynamics of two interacting
  particles, Phys. Rev. D 33 (1986) 3401--3424.
\newblock \href {https://doi.org/10.1103/PhysRevD.33.3401}
  {\path{doi:10.1103/PhysRevD.33.3401}}.

\bibitem{Petrat2014b}
S.~Petrat, R.~Tumulka, Multi-time equations, classical and quantum, Proceedings
  of the Royal Society A: Mathematical, Physical and Engineering Sciences
  470~(2164) (2014) 20130632.
\newblock \href {https://doi.org/10.1098/rspa.2013.0632}
  {\path{doi:10.1098/rspa.2013.0632}}.

\bibitem{Ptashynskiy2019arXiv}
D.~A. Ptashynskiy, T.~M. Zelentsova, N.~O. Chudak, K.~K. Merkotan, O.~S.
  Potiienko, V.~V. Voitenko, O.~D. Berezovskiy, V.~V. Opyatyuk, O.~V. Zharova,
  T.~V. Yushkevich, I.~V. Sharph, V.~D. Rusov, Multiparticle fields on the
  subset of simultaneity (2019).
\newblock \href {https://doi.org/10.48550/arXiv.1905.07233}
  {\path{doi:10.48550/arXiv.1905.07233}}.

\bibitem{Dicke1978}
R.~H. Dicke, {Is there a chronometer hidden deep in the sun}, Nature 276 (1978)
  676--680.
\newblock \href {https://doi.org/10.1038/276676b0}
  {\path{doi:10.1038/276676b0}}.

\bibitem{Dicke1979}
R.~H. {Dicke}, {Solar luminosity and the sunspot cycle}, Nature 280 (1979)
  24--27.
\newblock \href {https://doi.org/10.1038/280024a0}
  {\path{doi:10.1038/280024a0}}.

\bibitem{Dicke1988}
R.~H. Dicke, The phase variations of the solar cycle, Solar Physics 115~(1)
  (1988) 171--181.
\newblock \href {https://doi.org/10.1007/BF00146238}
  {\path{doi:10.1007/BF00146238}}.

\bibitem{Gould1987}
A.~{Gould}, {Resonant enhancements in weakly interacting massive particle
  capture by the earth}, Astrophysical Journal 321 (1987) 571--585.
\newblock \href {https://doi.org/10.1086/165653} {\path{doi:10.1086/165653}}.

\bibitem{LopesSilk2002}
I.~P. {Lopes}, J.~{Silk}, {Solar Neutrinos: Probing the Quasi-isothermal Solar
  Core Produced by Supersymmetric Dark Matter Particles}, Physical Review
  Letters 88~(15) (2002) 151303.
\newblock \href {http://arxiv.org/abs/astro-ph/0112390}
  {\path{arXiv:astro-ph/0112390}}, \href
  {https://doi.org/10.1103/PhysRevLett.88.151303}
  {\path{doi:10.1103/PhysRevLett.88.151303}}.

\bibitem{LopesSilk2010}
I.~{Lopes}, J.~{Silk}, {Neutrino Spectroscopy Can Probe the Dark Matter Content
  in the Sun}, Science 330 (2010) 462.
\newblock \href {https://doi.org/10.1126/science.1196564}
  {\path{doi:10.1126/science.1196564}}.

\bibitem{LopesSilk2012}
I.~{Lopes}, J.~{Silk}, {Solar Neutrino Physics: Sensitivity to Light Dark
  Matter Particles}, Astrophysical Journal 752 (2012) 129.
\newblock \href {http://arxiv.org/abs/1309.7573} {\path{arXiv:1309.7573}},
  \href {https://doi.org/10.1088/0004-637X/752/2/129}
  {\path{doi:10.1088/0004-637X/752/2/129}}.

\bibitem{Lopes2014}
I.~{Lopes}, P.~{Panci}, J.~{Silk}, {Helioseismology with Long-range Dark
  Matter-Baryon Interactions}, Astrophysical Journal 795 (2014) 162.
\newblock \href {http://arxiv.org/abs/1402.0682} {\path{arXiv:1402.0682}},
  \href {https://doi.org/10.1088/0004-637X/795/2/162}
  {\path{doi:10.1088/0004-637X/795/2/162}}.

\bibitem{Dikpati2008}
M.~Dikpati, P.~A. Gilman, G.~de~Toma, The waldmeier effect: An artifact of the
  definition of wolf sunspot number?, The Astrophysical Journal 673~(1) (2008)
  L99--L101.
\newblock \href {https://doi.org/10.1086/527360} {\path{doi:10.1086/527360}}.

\bibitem{Gough2010}
D.~Gough, Vainu bappu memorial lecture: What is a sunspot?, in: S.~Hasan, R.~J.
  Rutten (Eds.), Magnetic Coupling between the Interior and Atmosphere of the
  Sun, Astrophysics and Space Science Proceedings, Springer Berlin Heidelberg,
  2010, pp. 37--66.
\newblock \href {https://doi.org/10.1007/978-3-642-02859-5_4}
  {\path{doi:10.1007/978-3-642-02859-5_4}}.

\bibitem{Zioutas2009}
K.~Zioutas, M.~Tsagri, Y.~Semertzidis, T.~Papaevangelou, T.~Dafni,
  T.~Anastassopoulos, Axion searches with helioscopes and astrophysical
  signatures for axion(-like) particles, New J. Physics 11 (2009) 105020,
  \href{http://arxiv.org/abs/0903.1807}{arXiv:0903.1807 [astro-ph.SR]}.

\bibitem{Peres2000}
G.~Peres, S.~Orlando, Reale, R.~Rosner, H.~Hudson, The {Sun} as an {X}-ray star
  ii. using the {Yohkoh/Soft} {X}-ray telescope-derived solar emission measure
  versus temperature to interpret stellar {X}-ray observations, The Astrophys.
  J. 528 (2000) 537--551.

\bibitem{Gudel2004}
M.~G{\"u}del, X-ray astronomy of stellar coronae, The Astronomy and
  Astrophysics Review 12~(2) (2004) 71--237.
\newblock \href {https://doi.org/10.1007/s00159-004-0023-2}
  {\path{doi:10.1007/s00159-004-0023-2}}.

\bibitem{Paxton2010}
B.~Paxton, L.~Bildsten, A.~Dotter, F.~Herwig, P.~Lesaffre, F.~Timmes, Modules
  for experiments in stellar astrophysics (mesa), The Astrophysical Journal
  Supplement Series 192~(1) (2010) 3.
\newblock \href {https://doi.org/10.1088/0067-0049/192/1/3}
  {\path{doi:10.1088/0067-0049/192/1/3}}.

\bibitem{Gann2015}
G.~D.~O. Gann, Everything under the sun: A review of solar neutrinos, AIP
  Conference Proceedings 1666~(1) (2015) 090003.
\newblock \href {https://doi.org/10.1063/1.4915568}
  {\path{doi:10.1063/1.4915568}}.

\bibitem{Borexino2020a}
{Borexino Collaboration}, Sensitivity to neutrinos from the solar {CNO} cycle
  in {Borexino}, The European Physical Journal C 80~(11) (2020) 1091.
\newblock \href {https://doi.org/10.1140/epjc/s10052-020-08534-2}
  {\path{doi:10.1140/epjc/s10052-020-08534-2}}.

\bibitem{Borexino2020b}
{Borexino Collaboration}, Experimental evidence of neutrinos produced in the
  cno fusion cycle in the sun, Nature 587~(7835) (2020) 577--582.
\newblock \href {https://doi.org/10.1038/s41586-020-2934-0}
  {\path{doi:10.1038/s41586-020-2934-0}}.

\bibitem{TapiaArellano2021}
N.~Tapia-Arellano, S.~Horiuchi, Measuring solar neutrinos over gigayear
  timescales with paleo detectors, Physical Review D 103~(12) (2021) 123016.
\newblock \href {https://doi.org/10.1103/PhysRevD.103.123016}
  {\path{doi:10.1103/PhysRevD.103.123016}}.

\bibitem{Asplund2009}
M.~{Asplund}, N.~{Grevesse}, A.~J. {Sauval}, P.~{Scott}, {The Chemical
  Composition of the Sun}, Annual Review of Astron and Astrophys 47 (2009)
  481--522.
\newblock \href {http://arxiv.org/abs/0909.0948} {\path{arXiv:0909.0948}},
  \href {https://doi.org/10.1146/annurev.astro.46.060407.145222}
  {\path{doi:10.1146/annurev.astro.46.060407.145222}}.

\bibitem{Vincent2015a}
A.~C. {Vincent}, P.~{Scott}, A.~{Serenelli}, {Possible Indication of
  Momentum-Dependent Asymmetric Dark Matter in the Sun}, Physical Review
  Letters 114~(8) (2015) 081302.
\newblock \href {http://arxiv.org/abs/1411.6626} {\path{arXiv:1411.6626}},
  \href {https://doi.org/10.1103/PhysRevLett.114.081302}
  {\path{doi:10.1103/PhysRevLett.114.081302}}.

\bibitem{Vincent2015b}
A.~C. {Vincent}, A.~{Serenelli}, P.~{Scott}, {Generalised form factor dark
  matter in the Sun}, Journal of Cosmology and Astroparticle Physics 8 (2015)
  040.
\newblock \href {http://arxiv.org/abs/1504.04378} {\path{arXiv:1504.04378}},
  \href {https://doi.org/10.1088/1475-7516/2015/08/040}
  {\path{doi:10.1088/1475-7516/2015/08/040}}.

\end{thebibliography}

\end{document}